\documentclass{aa}  
\usepackage{graphicx}
\usepackage{txfonts}
\usepackage{natbib}

      \def\new#1 {{\bf #1 }}
      \def\cut#1 {\sout{#1} }
      
\usepackage{aalongtable}

\begin{document}
\title{
Coronae in the \textsl{\textbf{Coronet}}: \\ A very deep X-ray look into a stellar nursery
}

\author{Jan Forbrich\inst{1,2}\thanks{now at: Harvard-Smithsonian Center for Astrophysics, 60 Garden Street, MS 42, Cambridge, MA 02138, U.S.A.,  \email{jforbrich@cfa.harvard.edu}} \and Thomas Preibisch\inst{1}}

\institute{Max-Planck-Institut f\"ur Radioastronomie,
 Auf dem H\"ugel 69, D--53121 Bonn, Germany
 \and
 Astrophysikalisches Institut und Universit\"ats-Sternwarte Jena, Schillerg\"a{\ss}chen 2-3, D--07745 Jena, Germany
} 

   \date{Submitted: ; accepted: }

 
  \abstract
   {} 
   {To study the X-ray properties of young stellar objects (YSOs),
we analyze an exceptionally sensitive \textsl{Chandra} dataset
of the \textsl{Coronet} cluster in the CrA star-forming region, achieving a limiting 
luminosity of $L_{\rm X,min} \sim  5\times 10^{26}$~erg/sec for
lightly absorbed sources. This dataset represents one of the most
sensitive X-ray observations ever obtained of a star-forming
region.
}
   {
The X-ray data are used to investigate the membership 
status of tentative members of the region, 
to derive plasma temperatures and X-ray luminosities of the
YSOs, and to investigate variability 
on the timescale of several years.
}
   {46 of the 92 X-ray sources in the merged \textsl{Chandra} image
can be identified with optical or near/mid-infrared counterparts.
X-ray emission is detected from \mbox{\em all} of the previously known
optically visible late-type (spectral types G to M) stellar
cluster members, from five of the eight brown dwarf candidates, and
from nine
embedded objects (``protostars'') with class 0, class I, or flat-spectrum SEDs
in the field of view. 
While the Herbig Ae/Be stars 
TY~CrA and R~CrA, a close companion of the B9e star HD~176386,
and the F0e star T CrA are detected,
no X-ray emission is found from
any of the Herbig-Haro (HH)~objects or the protostellar cores without infrared
source. We find indications for diffuse X-ray 
 emission near R~CrA / IRS~7.
}
{The observed X-ray properties of the \textsl{Coronet} 
YSOs are consistent with coronal activity; no soft spectral components hinting
towards X-ray emission from accretion shocks were found.
The X-ray emission of the AeBe stars TY~CrA and HD~176386  
originates probably from close late-type companions.
The Ae star R~Cra shows a peculiar X-ray spectrum and
an extremely hot plasma temperature.
Finally, we discuss the differences of the X-ray properties of YSOs in 
different evolutionary stages.
}

\titlerunning{Coronae in the \textsl{Coronet}
}

   \maketitle


\section{Introduction}

Young stellar objects (YSOs) generally show high levels 
of X-ray activity, exceeding the emission level of the Sun 
and late-type field stars by several orders of magnitudes
\citep[for a recent review see][]{Feigelson06}.
A good knowledge of the X-ray properties of YSOs
is of paramount importance not
only for the understanding of the physical mechanisms that lead to
the X-ray emission; the X-ray emission
has also far-reaching implications
for the physical processes in the circumstellar environment, the
formation of planetary systems, and the evolution of 
protoplanetary atmospheres
\citep[e.g.,][]{Glassgold05,Feigelson05b}.
In the investigation of the stellar populations of 
star-forming regions,
X-ray studies are particularly effective in discriminating
YSOs from unrelated fore- and background field stars.
X-ray studies can give a census of the members of a star-forming
region that is independent of circumstellar material, allowing
to overcome the bias in membership determinations based on
infrared excess criteria.
Furthermore, since radiation at energies above $\sim 1$~keV
is much less affected by extinction than optical light,
X-ray observations can penetrate up to $A_V \sim 500$~mag into the 
cloud and allow a deep look at embedded YSOs.

Recently, two very large observational projects provided unprecedented
X-ray datasets on young stars.
The $Chandra$ Orion Ultradeep Project (COUP), 
a 10-day long 
observation of the Orion Nebula Cluster 
with $Chandra$/ACIS \citep[see][]{Getman05a} is the
deepest and longest X-ray 
observation ever made of a young stellar cluster.
With a detection limit of 
$L_{\rm X,min} \sim 10^{27.3}$~erg/sec for lightly absorbed 
sources, X-ray emission from more than 97\% of the 
$\sim 600$ 
optically visible and well characterized late-type 
(spectral types F to M) cluster stars
was detected \citep{Preibisch_coup_orig}.
The 
XMM-Newton Extended Survey of the Taurus Molecular Cloud (XEST)
covered the densest stellar populations in a 5 square degree region
of the Taurus Molecular Cloud \citep[see][]{Guedel07}
and provided X-ray data on 110 optically well characterized young stars.
Despite the new dimension in the quantity and quality of the
X-ray data on these two star-forming regions,
sensitive X-ray observations of other star-forming regions
are still useful to shed new or additional light on several 
still open questions.

The first question concerns the origin of the X-ray emission in T Tauri stars
(TTS).
The hot ($\ga 10 - 30$~MK) plasma temperatures typically derived from
the X-ray spectra of TTS show quite clearly that the 
bulk of the X-ray emission must be related to coronal 
magnetic activity \citep{Preibisch_coup_orig}, a conclusion that is also
supported by the lack of correlated X-ray and optical variability
\citep{Stassun06,Stassun07,Forbrich06b}.
However, in some TTS, some fraction of the X-ray emission seems to
originate in accretion shocks at the
stellar surface (e.g. \citealp{Kastner02}) 
and/or in shocks in the innermost parts of stellar jets 
(e.g. \citealp{Guedel07a}).
Due to the relatively low shock temperatures of at most a few MK,
such shock-related X-ray emission should 
be detected as a soft excess (at energies below $\sim 1$~keV)
superposed onto the much harder coronal emission
(e.g. \citealp{Schmitt05,Guenther07}). Although high-resolution X-ray spectra
are required for a detailed investigation of the origin of the
different spectral components \citep[e.g.][]{Telleschi07a},
in some cases indications for very soft spectral components have also
been found in CCD low-resolution spectra of young stars
(e.g., \citealp{Flaccomio06}; see also the model spectra
in \citealp{Guenther07}).

A related question concerns the
origin of the observed X-ray emission from
young intermediate-mass (Herbig Ae/Be) stars.
As these intermediate-mass stars
have neither outer convection zones that may harbor a dynamo
to produce magnetic activity, nor strong enough stellar winds
to create X-rays in internal wind shocks,
the detection of X-ray emission from a large fraction
of the observed Herbig Ae/Be stars  still 
remains largely
unexplained \citep[e.g.,][]{dam94,zip94,Hamaguchi05a,Stelzer05,Stelzer06}.
\textsl{Chandra} observations with their superior spatial resolution
revealed that in some of the X-ray--detected Herbig Ae/Be stars 
the true source of the X-ray
emission is not the A or B star, but a nearby late-type companion.
It remains unclear, however, whether late-type companions are the true
source of the X-ray emission in {\em all} cases, or whether some
Herbig Ae/Be stars may nevertheless be intrinsic X-ray emitters.
Some Herbig Ae/Be stars, e.g. AB Aur and HD~163296,
show very soft X-ray spectra that have been interpreted
as emission from magnetically confined
winds \citep{Telleschi07b} or accretion shocks \citet{Swartz05}.
Obtaining good S/N X-ray spectra of further Herbig Ae/Be stars can help to
investigate these possibilities.

Another open question is how
early in the protostellar evolution X-ray
activity start.
While class I protostars
are well established X-ray emitters
\citep[e.g.,][]{Grosso97,Neuhaeuser97,Imanishi01a},
it is still unclear whether
class~0 protostars, which represent an even earlier,
extremely young evolutionary stage in which most of the
mass resides still in the circumstellar environment,
also show X-ray activity.
The detection of an X-ray flare from the candidate
class~0 protostar IRS~7E in the \textsl{Coronet} cluster
by \citet{ham05b}  provided the first piece of evidence for the
presence of X-ray emission in extremely young objects,
but the exact evolutionary stage of IRS~7E is not yet fully clear.
This object clearly deserves further examination.
Having no near-infrared counterpart, this source was recently
classified as a class~0/I transitional object by \citet{Groppi07}, 
based on mid-infrared detections and new submillimeter data.

A final interesting issue is X-ray emission from HH objects.
Since the X-ray detection of 
HH~2 \citep{Pravdo01}, it is clear that 
the shock-heated material in {\em some} jets
can actually produce observable soft X-ray emission 
\citep[e.g.][]{Pravdo04,Grosso06,Favata06}.
 However, the vast majority of
all HH~objects remain undetected in X-ray observations.
It is not clear whether this is due to the limited sensitivity 
of existing X-ray observations, or whether X-ray emission
is created only in a small fraction (the fastest ?) of all
jets.
Very deep X-ray data of nearby star-forming regions allow to
investigate this point.

The deep \textsl{Chandra} data discussed in this
paper are well suited to address all these issues.
The Corona Australis star formation region 
is one of the nearest (about 3.5 times closer than the Orion Nebula
Cluster) and most 
active regions of recent and ongoing star formation 
\citep[e.g.][]{Neuhaeuser00,NF07}.
It contains a loose cluster of a few dozen known YSOs, which
cover a wide range of masses (from intermediate-mass Herbig AeBe 
stars down to very-low-mass brown dwarfs) and evolutionary stages
(from
pre-stellar cores through class~0 and class~I protostars,
class~II T Tauri stars, to class~III objects
that have already cleared their dusty environment).
The central part of the star-forming region,
around the bright star R~CrA, contains the densest clustering of
very young, embedded objects, which is known as the ``\textsl{Coronet}''
cluster. 
Dozens of HH objects trace jets emanating
from the YSOs. We refer to the region covered by the X-ray data discussed here as the \textsl{Coronet} region (see Fig.~\ref{spitzer_cxs}) because it is centered on the \textsl{Coronet} cluster and covers its surroundings.
The distance to the Corona Australis star-forming region is relatively well known,
based on distance determination of two members.
The optically brightest member, the B8e star TY~CrA, is a well known eclipsing
spectroscopic multiple system, for which 
\citet{Casey98} derived a distance of
$D = 129 \pm 11$~pc. The B9e star HD~176386,
which forms a common proper motion pair with TY~CrA 
\citep{Teixeira00} and shows signs of strong accretion \citep{Grady93},
has a Hipparcos distance of $136^{+25}_{-19}$~pc, which
 is fully consistent with the distance derived for TY~CrA.
We thus adopt a distance of 130~pc for the CrA star-forming region.

The CrA star-forming region was the target of numerous X-ray observations
prior to the present study, including observations with 
EINSTEIN \citep{dam94,Walter97},
ROSAT \citep{zip94,Neuhaeuser97,Neuhaeuser00},
ASCA \citep{koy96},
\textsl{Chandra} \citep{gar03},
and
XMM-\textsl{Newton} \citep[see][]{ham05b,Forbrich06a}.
All these observations, however, were several times less sensitive than the
dataset analyzed in the present paper.


\section{{\em Chandra} observations and data analysis}

We have performed a series of five individual $\sim 15$~ksec 
observations
(separated by about one day)
of the \textsl{Coronet} region with the ACIS 
camera on-board \textsl{Chandra} \citep{Weisskopf02,Garmire03}.
The main aim of these observations was to monitor the X-ray 
emission of the YSOs simultaneously with optical, infrared,
and radio observations, and the results of this multi-wavelength
monitoring are described in \cite{Forbrich06b}.
 In order to 
optimize the sensitivity of the X-ray data for the present study,
we included into our analysis three previous \textsl{Chandra}
observations retrieved from the public archive, which have exposure
times between 20 and 40~ksec. Details of the individual observations
are listed in Table~\ref{table_obs}.

\begin{table}
\centering
\caption{\textsl{Chandra} observations of the CrA star-forming region used
in this study. 
}
\label{table_obs}       
\begin{tabular}{rccc}
\hline\noalign{\smallskip}
Obs & Expo. & Aimpoint & Start Date \\
 & [ksec] &  [J2000] & \\
\noalign{\smallskip}\hline\noalign{\smallskip}
 19   & 19.96 & 19 01 50.6 \, $-36$ 57 30 &  2000-10-07 \, 17:00:55\\
3499  & 38.13 & 19 01 50.6 \, $-36$ 57 30 &  2003-06-26 \, 12:57:06\\
4475  & 20.17 & 19 01 48.9 \, $-36$ 59 23 &  2004-06-17 \, 23:15:56\\
5402  & 15.39 & 19 01 45.0 \, $-36$ 58 09 &  2005-08-08 \, 02:36:49\\
5403  & 15.15 & 19 01 45.0 \, $-36$ 58 09 &  2005-08-09 \, 02:37:49\\
5404  & 15.17 & 19 01 45.0 \, $-36$ 58 09 &  2005-08-10 \, 01:57:21\\
5405  & 15.17 & 19 01 45.0 \, $-36$ 58 09 &  2005-08-12 \, 03:11:53\\
5406  & 14.76 & 19 01 45.0 \, $-36$ 58 09 &  2005-08-13 \, 01:50:17\\
\noalign{\smallskip}\hline
\end{tabular}
\end{table}

Our data analysis was performed with the
Chandra Interactive Analysis of Observations (CIAO) software 
package version 3.3, combined with CALDB 3.2.1.
Merging all eight individual observations results in a very deep 
dataset with a total exposure time of 156\,470~sec.
In our analysis we consider only
results arising from the imaging array (ACIS-I) of four abutted
$1024 \times 1024$ pixel front-side illuminated charge-coupled
devices (CCDs) providing a field of view of
 about $17\arcmin \times 17 \arcmin$ on the
sky. As the aimpoints and the orientation of the camera on the
sky (the ``roll angle'') differed for the
individual observations, the merged image
covers a slightly larger area.
Exposure maps and aspect histograms were computed for the single datasets as well as for the merged dataset,
allowing for a determination of the effective exposure time at each
sky position,  taking into account the spatial variation of
the detector
quantum efficiency, non-uniformities across the face of a
detector, mirror vignetting, and bad pixels.

The detection limit of the \textsl{Chandra} data was determined in
the following way:
using the PIMMS\footnote{PIMMS is the
Portable, Interactive Multi-Mission Simulator provided by the
HEASARC Online Service at
http://heasarc.gsfc.nasa.gov/Tools/w3pimms.html}
 software developed by the NASA High
 Energy Astrophysics Science Archive Science Center
and assuming an intrinsic source 
spectrum of a 10\,MK thermal plasma with a metal abundance 
of 0.4 times solar, as
typical for young stellar X-ray sources \citep[e.g.,][]{Getman05a},
we find that 
one detected count in 156.47~ksec corresponds to
an X-ray flux of $4.2\times 10^{-17}\,{\rm ergs/cm^2/s}$ 
in the $0.2\!-\!8$~keV energy range.
A five count source detection thus corresponds to an X-ray luminosity
of $L_{\rm X,min} \sim 4.3\times 10^{26}$~erg/s at the assumed distance 
of 130~pc and for no intervening extinction.
Our dataset thus represents one of the most sensitive
X-ray observations of a star-forming region ever obtained; 
it is about four times more sensitive than the data of
the \textsl{Chandra} Orion Ultradeep Project \citep{Getman05a},
which provided a detection limit of 
$L_{\rm X,min} \sim 2 \times 10^{27}$~erg/sec for lightly absorbed 
sources \citep{Preibisch_coup_orig}. 
Many of the YSOs in the CrA star-forming region
suffer from substantial extinction, up to $A_V \sim 45$\,mag.
The intervening extinction raises the effective detection limit;
for $A_V = [3,  10, 20, 45]$~mag the 
detection limits are $[1.4, 6.5, 19, 58] \times 10^{27}$~erg/s.
Note that these values are valid for the central ($\sim 5'$ radius) 
area of the \textsl{Chandra} image;
source detectability degrades with off-axis angle,
and is $\sim 2-3$ times lower near the edges of the \textsl{Chandra} 
field of view (at offaxis angles $8' - 11'$).

For locating the X-ray sources in our image, we
used the wavelet transform detection algorithm implemented as
the {\rm{\it wavdetect}} program within the CIAO data analysis
system \citep{Freeman02,Getman05a}.

With a nominal threshold of identifying a pixel as belonging to a source (parameter 'sigthresh') of $1.6 \times 10^{-7}$ and wavelet scales between 
1 and 16 pixels, the program located 91 sources. We manually added one clearly
detected source with about 100 counts (marked as such in Table~\ref{acis_extract.tab}),  which was missed by the algorithm
due to its location very close to the edge of the image. Thus, 
we consider a total of 92 sources.


After reprocessing all observations with CIAO 3.3 yielded a homogeneous set
of ''level 2'' event files (without pixel randomization), the {\rm{\it acis\_extract}} package\footnote{see
{http://www.astro.psu.edu/xray/docs/TARA/code/}}, version 3.94, was used for further analysis. Spectral extraction regions were 
defined in a way to include a specified
fraction of the point-spread function at the respective positions (90\% at an energy of 1.5~keV), independently for each observation.
In the framework of {\rm{\it acis\_extract}}, the background was determined from a region surrounding each source, containing a minimum number of counts (100 in our case) and excluding neighbouring sources.
Composite source spectra were constructed by summing the 
single-observation spectra, taking into account appropriately scaled 
background spectra for each observation. Composite response matrix 
files (RMFs) and ancillary response files (ARFs) were constructed 
using the FTOOLs \textsl{addrmf} and \textsl{addarf}, weighting the 
single-observation source-specific RMFs and ARFs by their respective 
exposure times.  
Spectral fitting with Monte Carlo techniques was then performed with the CIAO tool \textsl{Sherpa} (see Sect.~\ref{sect.xspec}).

Basic results of this analysis, namely a source list with net counts, median energy, source significance, hardness ratios, as well as incident flux estimates and information on the effective exposure times are listed in Table~\ref{acis_extract.tab}. The errors for the net counts contain the propagated $1\sigma$ Gehrels errors \citep{Gehrels86} of the source and background counts. The uncertainty of the absolute energy calibration of ACIS is 0.3\%\footnote{see http://cxc.harvard.edu/cal/}. Also listed is the source significance, i.e. the photometric
signal-to-noise ratio. Hardness ratios compare the source counts in two energy bands in the form HR = (Cts$_{\rm hard}$ - Cts$_{\rm soft}$)/(Cts$_{\rm hard}$ + Cts$_{\rm soft}$). We list the three hardness ratios as defined in 
\citet{Getman05a} with the following energy ranges:  [0.5-2.0]~keV vs.~[2.0-8.0]~keV for HR1, [0.5-1.7]~keV vs.~[1.7-2.8]~keV for HR2, and
[1.7-2.8]~keV vs.~[2.8-8.0]~keV for HR3. 
The uncertainties for the hardness ratios were calculated using a method described by \citet{Lyons91} with a script supplementary to \rm{\it acis\_extract} developed by Konstantin Getman\footnote{http://www.astro.psu.edu/users/gkosta/AE/accessory\_tools.htm}. 
In a few cases, no error could be estimated because divisions by zero occur, cases correspondingly marked by 'NaN' (for 'not a number'). Finally, we list the two \rm{\it acis\_extract} flux estimates. Their difference lies in the handling of the ancillary response matrix which is done either channel-wise (for FLUX1) or using an averaged value (for FLUX2). For a detailed discussion of the algorithms used by \rm{\it acis\_extract}, see \citet{Getman05a} and the online documentation\footnote{http://www.astro.psu.edu/xray/docs/TARA/ae\_users\_guide.html}.

For those (putative) members of the CrA star-forming region that 
remained undetected
in the \textsl{Chandra} data, we
determined upper limits to the count rates and X-ray luminosities
in the following way. We 
counted the observed number of photons in source regions centered
at their optical/infrared positions and compared them to the expected
number of background photons determined from several large source-free
background regions.
We used the Bayesian statistics method described by  \cite{Kraft91}
to determine the 90\% confidence upper limits for their count rates.
From these count rate upper limits we computed upper limits for the
extinction corrected  X-ray luminosities in the $0.2\!-\!8$~keV 
band assuming thermal plasma-spectra  with a temperature
of $10$~MK 
and computing the absorbing hydrogen column density from
the visual extinction according to the empirical relation
$ N_{\rm H} \sim A_V\,\cdot\,2 \times 10^{21}\,\rm cm^{-2}$
\citep{Feigelson05a}.

\section{X-ray sources and cluster members}

\subsection{Source identification}

\begin{figure*}
\begin{center}\includegraphics*[width=16cm,bb= 25 197 568 656]{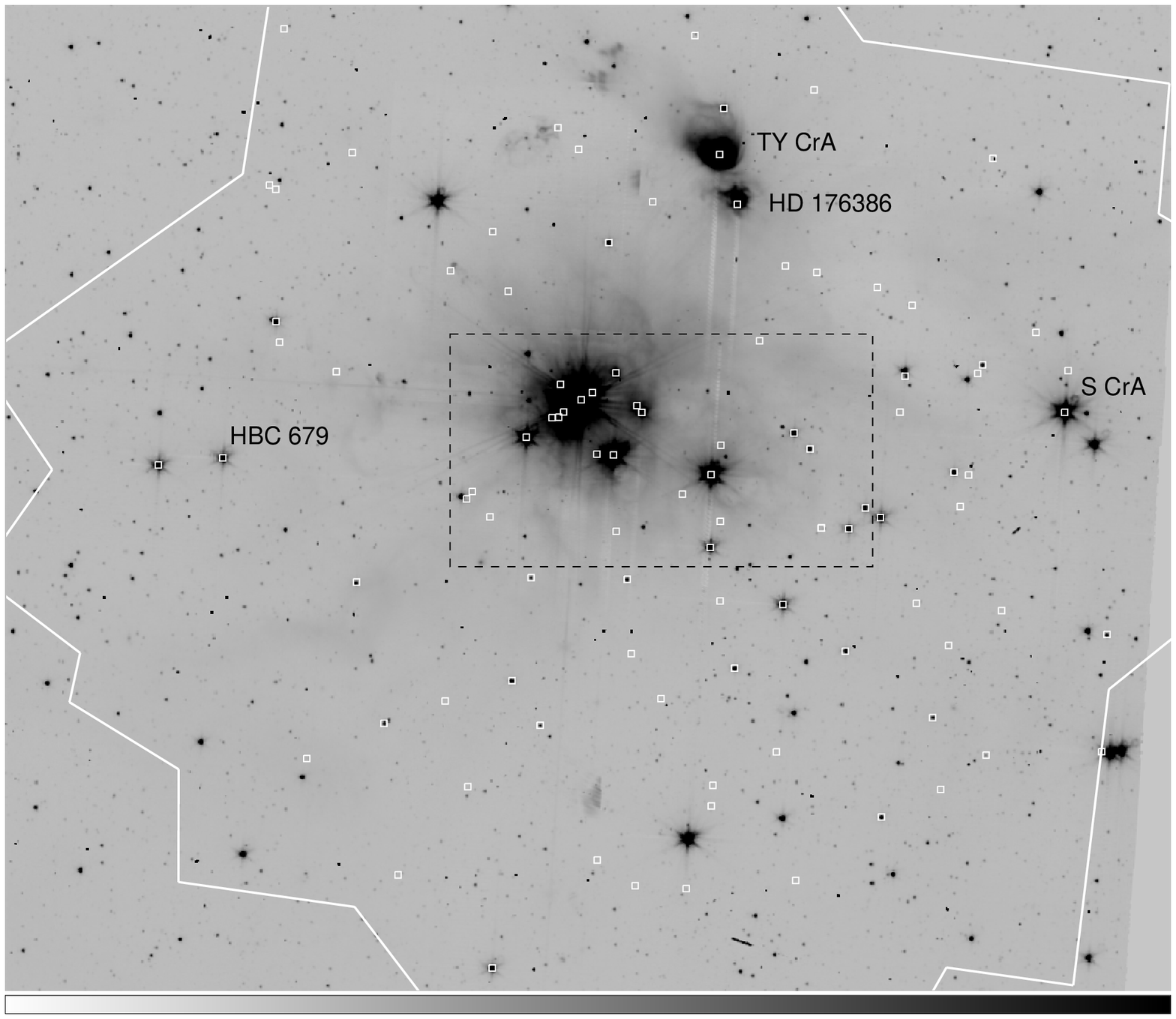}
\includegraphics*[width=10cm,bb= 25 280 568 577]{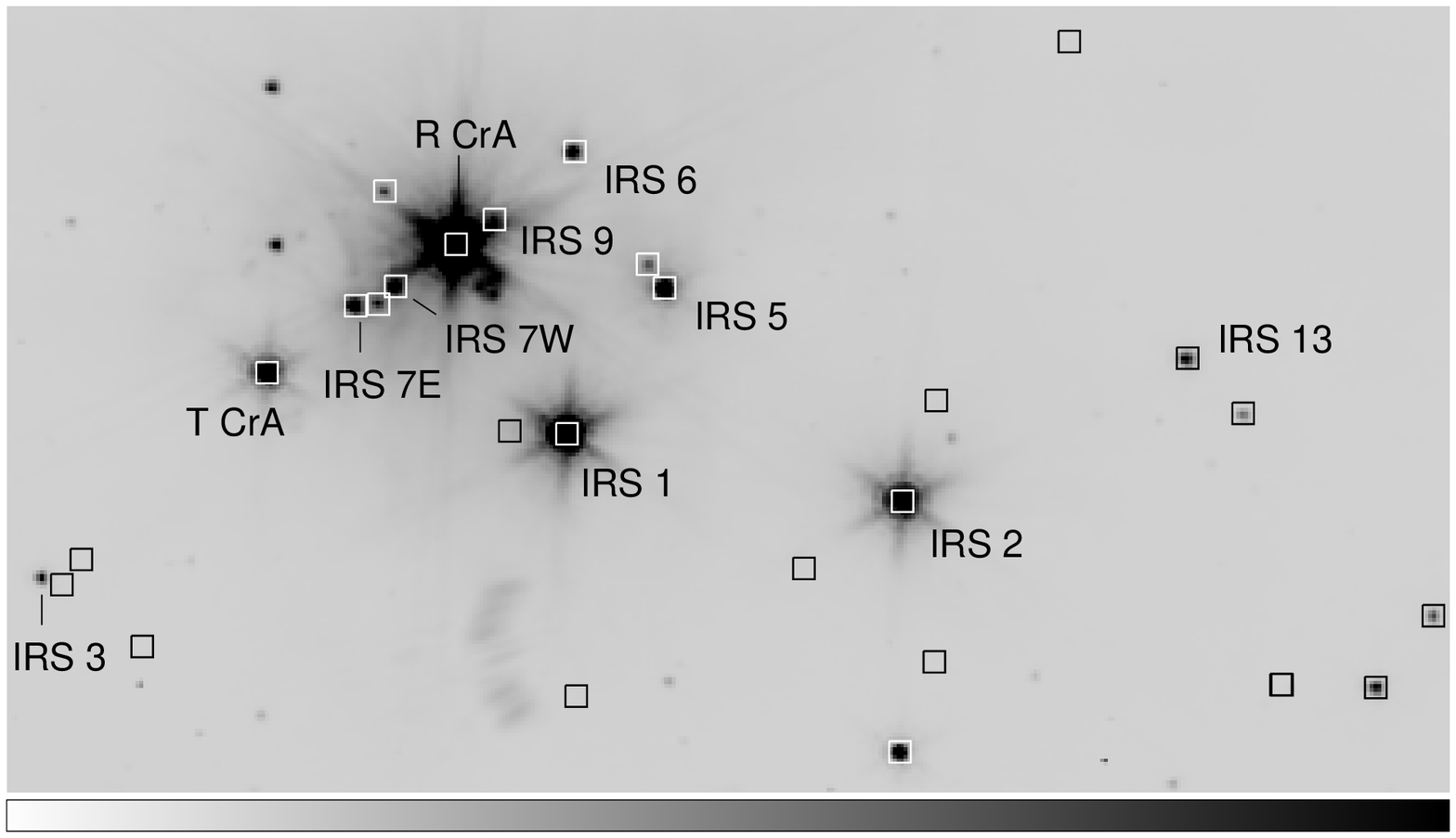}\end{center}
\caption{Above: $21.8' \times 18.4'$ detail of the
\textsl{Spitzer} IRAC $3.6\,\mu$m image of the \textsl{Coronet} region. The 
white line mark the boundaries of the field of view
of the merged \textsl{Chandra} data.
The positions of the \textsl{Chandra} X-ray sources are marked by the
boxes. The dashed black line indicates the area shown in the image
below.
Below: The central $7.8' \times 3.2'$ region of the \textsl{Spitzer} 
IRAC $4.5\,\mu$m image with
X-ray source positions marked by boxes.
} \label{spitzer_cxs}%
\end{figure*}

\begin{table*}[!ht]
\caption{Identification of \textsl{Chandra} X-ray sources with optical 
or infrared counterparts in the CrA star-forming 
region. Columns 2-4 provide information on counterparts to the
X-ray sources in the optical DSS images, the 2MASS images,
and the \textsl{Spitzer} images: "y" means that a counterpart exists, "n"
means that no counterpart can be seen. Column 5 gives the SED class 
derived from the Spitzer photometry, and the last 3 columns give
names, spectral types, and stellar bolometric luminosities of the stars
collected from
\citet{Wilking92,Walter97,Casey98,Olofsson99,Neuhaeuser00,Prato03,Nisini05};
additional references are given in the discussion of the individual
objects in Sections 3 and 4. }
\label{source_id_tab}       
%
\begin{tabular}{lccclllr}
\hline\noalign{\smallskip}
 Source     &\multicolumn{3}{c}{Counterpart}&SED &  Name & SpT & $L_{\rm bol}$  \\
  CXO J          & DSS & 2MASS &\textsl{Spitzer}& class &     &   & $[L_\odot]$ \\ \hline
190104.58--370129.6 &n & y& y& III& CrA~453   & M4 &  \\
190108.60--365721.3 &y & y& y& II & S CrA     & K3 & 2.29   \\
190115.86--370344.3 &n & n& y&    &           &    &   \\
190116.26--365628.4 &y & y& y& II & V667 CrA  &  M5  & \\ 
190118.90--365828.4 &n & y& y& II & CrA 466   & M4.5 &   \\
190119.39--370142.0 &n & n& y&    &           &      & \\
190120.86--370302.9 &y & y& y& III& CrA 4111  & M5   & \\
190122.40--370055.4 &n & n& y&    &           &      &   \\
190125.61--370453.9 &n & y& y& III& ISO-CrA 133   &  & \\ 
190125.75--365919.3 &n & y& y& II & ISO-CrA 134   &  & \\
190127.15--365908.6 &y & y& y& III& ISO-CrA 135   & M & 0.12 \\
190128.72--365931.9 &y & y& y&    & ISO-CrA 136   &  & \\
190129.01--370148.8 &y & y& y& III& ISO-CrA 137   &  & \\
190132.34--365803.1 &n & y& y& II &TS 2.9 = ISO-CrA 139 &  & \\   
190133.84--365745.0 &n & y& y& II &TS 2.8 =  IRS 13   &  &0.07  \\
190134.84--370056.7 &y & y& y& III& V709 CrA    &  K1\,IV & 1.55  \\  
190139.15--365329.6 &y & y& y&    & HD 176386-B  &  & \\
190139.34--370207.8 &y & y& y& III&          &   & \\
190140.40--365142.4 &n & y& y& II &          &   & \\
190140.81--365234.0 &y & y& y&    & TY CrA\, a/b/c/d  &  B8e/K2/F--K/M  &   \\
190141.55--365831.6 &n & y& y& I  & IRS 2    &  K2  & 4.3 \\
190141.62--365953.1 &y & y& y& II & HBC 677  &   & \\
190143.12--365020.9 &n & n& y&    &          &   & \\
190148.02--365722.4$^1$&n & y& y& I  & IRS 5 a/b  & K6\,V/?  & 1.6/? \\
190148.46--365714.5 &n & n& y& I  &          &   & \\
190149.35--370028.6 &y & y& y& III& LS-RCrA 2& M6 (BD cand.) &  \\
190150.45--365638.1 &n & y& y& II & IRS 6A   & M2  & 0.5 \\
190150.66--365809.9 &n & y& y& I  & V710 CrA  = IRS 1 = HH 100 IR & K5-M0 & 3.1 \\
190151.11--365412.5 &n & y& y& II & IRS 8    &  & \\
190152.63--365700.2 &n & y& y& I  & IRS 9    &  & \\
190153.67--365708.3 &y & y& y&    & R CrA    & A5\,IIe\,var  & \\
190155.31--365722.0 &n & n& y& I  & IRS 7\,W (= 7\,A)  &  &  \\
190155.61--365651.1 &n & n& y& I  &          &  & \\
190155.76--365727.7 &n & n& y& I/II&         &  & \\
190155.85--365204.3 &n & n& y&    &          &  &   \\
190156.39--365728.4 &n & n& y& 0/I& IRS 7\,E  (= 7\,B) &  & \\
190157.46--370311.9 &n & y& y&    &          &  & \\
190158.32--370027.5 &n & n& y& III&          &  & \\
190158.79--365750.1 &y & y& y& II & T CrA    & F0e  & 3.4\\ 
190200.11--370222.3 &y & y& y& III& 2MASS J19020012-3702220 & M4  & \\
190201.92--370743.0 &y & y& y& III& V702 CrA & G5  & 1.55\\
190201.94--365400.1 &n & n& y&    & B185839.6-365823  &  BD cand.  & \\
190211.99--370309.4 &n & y& y& II & ISO-CrA 143  &  BD cand. & 0.01 \\ 
190222.13--365541.0 &y & y& y& III& HBC 679  & K2\,IV  & 0.50 \\ 
190227.05--365813.2 &y & y& y& II & HBC 680  &M  & 0.72 \\ 
190233.07--365821.1 &y & y& y& II & ISO-CrA 159  & M & 0.45  \\
\noalign{\smallskip}\hline
\end{tabular}\smallskip

$^1$ The X-ray emission of the $0.9''$ binary IRS 5a/b \citep{Chen93,Nisini05}
is marginally resolved in the \textsl{Chandra} image. 
\end{table*}

In order to identify the X-ray sources, we
inspected the source positions on optical images
from the Digitized Sky Survey, near-infrared images from 
2MASS, and near- to mid-infrared images from the \textsl{Spitzer}
observatory. 

A set of reduced \textsl{Spitzer}-IRAC mosaic images of the CrA star-forming region and a list of sources with IRAC photometry 
and classifications of their broad-band infrared spectral energy
distributions (SEDs)
was kindly provided to us by Lori Allen.
The SED classification was performed as described in \cite{Megeath04}
and sorts the objects into the different infrared classes 0/I/II/III
\citep[with class~0 = protostar, class~I =
evolved protostar, class~II = T Tauri star with disk, class~III =
T Tauri star without disk; for further information 
see][]{Lada87,Andre93}.
The superb PSF of \textsl{Chandra}/ACIS and the high
accuracy of the aspect solution, resulting in a positional
accuracy of typically better than $1''$, allowed a clear and
unambiguous identification of 46 of the 92 X-ray sources
with optical and/or infrared counterparts. Figure~\ref{spitzer_cxs}
shows the location of the X-ray sources marked on the
$3.6\,\mu$m and the $4.5\,\mu$m \textsl{Spitzer} images.
Further information about the counterparts was obtained from
the SIMBAD database and the literature; references are given in the
text describing the individual objects.
The results of the source identification
are listed in Table \ref{source_id_tab}.

46 X-ray sources have either no counterparts in any of the
optical/infrared images mentioned above.
Given the detection limits of
$K_{\rm lim} \sim 15$ for the 2MASS image and 
our estimate of $L_{\rm lim} \ga 15$
for the \textsl{Spitzer} $3.6\,\mu$m image, it
is very unlikely that these objects are related to the
CrA star-forming region:
According to the \cite{Siess00} pre-main sequence (PMS) models,
a 5~Myr old $0.1\,M_\odot$ star at a distance of 130~pc 
would have un-reddened magnitudes of
$K = 9.5$ and $L = 9.2$. 
An extinction of $A_V \sim 50$~mag would thus 
be required to prevent detection
in the 2MASS $K$-band image, and an even higher value of
$A_V \sim 100$~mag to prevent detection in the \textsl{Spitzer} $3.6\,\mu$m image.
We also note that the
number of X-ray sources without counterparts
is in good agreement with the
expected number of (mostly extragalactic)
background X-ray sources that has been derived from the deep 
X-ray source counts\footnote{Based on the X-ray source counts 
presented by \citet{Brandt01} there should be
$\sim 60\!-\!70$ extragalactic sources exceeding
the flux limit of our data. The number of actually detectable 
extragalactic sources must be somewhat smaller 
due to the extinction 
of the dark cloud, especially in the central part of our field of view.
}.  
Furthermore, the background nature of these sources is also 
supported by 
their uniform spatial distribution within the field of view.

We also searched for, but could not find any X-ray emission 
associated
with the protostellar cores in the CrA star-forming region as listed
by \citet{Nutter05}, unless these contain infrared objects.
Finally, we also searched for X-ray sources at the positions
of 43 HH objects in the CrA star-forming region as listed 
in \citet{Wang04}.
From  none of these objects X-ray emission is detected\footnote{
We note that the \textsl{Chandra} source 190205.84--365444.2
has neither optical nor infrared counterparts, but
is located near a bow shock seen in the
\textsl{Spitzer} images.
However, a relation between the X-ray emission
and the bow shock seems very unlikely. First, the positional
offset of $8''$ is clearly much larger than the astrometric
accuracy, and the offset is {\em not} along the 
flow direction, i.e.~cannot be explained by the non-simultaneous 
nature of the X-ray and infrared images. 
Second, the distribution of energies
of the 21 detected source counts is quite hard (no photons
with energy $\leq 1$~keV, 4 photons in the $1-2$~keV, and 
17 photons in the  $2-8$~keV range), in strong contrast to the
very soft X-ray emission expected from jet shocks and observed 
in some cases \citep[e.g.][]{Pravdo04,Grosso06}. We therefore believe that the
X-ray emission comes from a background object and is not related
to the bow shock.}.

As discussed in the following subsections in more detail, 
we detect X-ray emission from
9 class~0, class~I, and flat-spectrum objects,
14 class~II objects, and 12 class~III objects.


\subsection{X-ray emission and membership in the star-forming region}

The observed X-ray properties can provide crucial information
for a clarification of
the membership status of presumed members discussed in the literature.
We can confidently 
expect to detect X-ray emission from (almost) any stellar member
of the region, given  the following considerations:
The COUP data showed that {\em all} young late-type stars in the Orion Nebula Cluster 
show strongly
elevated X-ray emission as compared to the Sun and solar-like
field stars \citep{Preibisch_coup_orig}: 
 at least 98\% of all late type YSOs 
in the Orion Nebula Cluster have fractional X-ray luminosities of
$\log\left(L_{\rm X}/L_{\rm bol}\right) > -5$, and there are
strong indications that the
 2\% of the stars below this value are not cluster members but field
stars \citep[see discussion in][]{Preibisch_coup_orig}. 

Assuming a lower limit to the fractional X-ray luminosities
of $\log\left(L_{\rm X}/L_{\rm bol}\right) > -5$
to hold for the YSOs in the CrA star-forming region,
we expect to detect X-ray emission from 
any of its stellar members, unless the extinction is too high. 
To put this statement in a more quantitative way, 
we conservatively assume that the YSOs have fractional X-ray 
luminosities of
$\log\left(L_{\rm X}/L_{\rm bol}\right) = -5$, and ages
of 5~Myr.
For the case of no extinction, the
X-ray detections should then be 100\% complete down to stars of
$\sim 0.08\,M_\odot$ or spectral type $\sim$~M7.
For an extinction of $A_V = 5$~mag, the limit of
complete detections is 
at $\sim 0.2\,M_\odot$ or spectral type $\sim$~M5, 
and 
for $A_V = 10$~mag at $\sim 0.5\,M_\odot$ 
or spectral type $\sim$~M0.
Note that the {\em typical} X-ray emission level of young stars is 
$\log\left(L_{\rm X}/L_{\rm bol}\right) =  -3.6$,
i.e.~a factor of 25 higher than
our assumed lower limit, so we expect to detect also the majority
of lower-mass objects, unless their extinction is very high.

\subsubsection{Previously known and suspected members}

\citet{Neuhaeuser00}
  list 34 previously known or suspected members of the CrA star-forming region.
Among the objects in the field of view of the \textsl{Chandra} data,
only two remain undetected.
2MASS~J19010586-3657570 (= ISO~CrA~113, spectral type G0)
shows  no near-infrared excess
and the \textsl{Spitzer} photometry is consistent with purely photospheric 
emission.
\citet{Neuhaeuser00} reported that no signs of 6708\,\AA\, Lithium absorption
are seen in its spectrum, suggesting that it is not a young star.
The second case is 2MASS J19014791-3659302 (= TS~13.4), which also
shows no infrared excess and has a class~III \textsl{Spitzer} SED.
Together with the non-detection in the \textsl{Chandra} data, these points
strongly suggest that these two stars are not members of
the CrA star-forming region but unrelated field stars.

The results of a ISOCAM survey of the CrA star-forming region 
were reported by \citet{Olofsson99}. They identified 21 infrared
sources with mid-infrared excess, 10 of which are located within
the field of view of the \textsl{Chandra} data. Eight of these 10
objects are detected as X-ray sources, only 
ISO~CrA~140 and ISO~CrA~145 remain undetected by \textsl{wavdetect}.
The 2MASS colors of ISO~CrA~140 show a near-infrared excess, but the
\textsl{Spitzer} photometry suggest a class~III SED, and we note that this
object was not detected in the ISOCAM $14.3\,\mu$m band, raising doubts
about the presence of excess emission.
The near-infrared colors of this object suggest an extinction on the order of
$A_V \approx 10$~mag.
Since ISO~CrA~140 is a relatively faint infrared source,  we
suspect that its low intrinsic luminosity in
combination with strong extinction may have prevented the X-ray
detection.
Alternatively, it may be an unrelated background AGB star 
rather than a young star in the CrA star-forming region.
With 28 counts in an area of radius $3''$ ($18.5^{+5.9}_{-4.8}$ net counts), ISO~CrA~145 may actually be marginally detected although the source was not found by \textsl{wavdetect}. The same region for ISO~CrA~140 contains only 10 counts.

\citet{Lopez05} identified 13 candidate very low-mass members of the
CrA star-forming region by optical spectroscopy. 
Six of these are in the field of view of the \textsl{Chandra} image,
and five of them, CrA~453, 466,
468, 4110, and 4111, all with estimated spectral types
around M5,
are detected as X-ray sources, consistent with the assumption
that they are young members of the CrA star-forming region.
The X-ray detection of all these very low-mass star reinforces our
expectation that our X-ray data should be complete for all stellar
members of the \textsl{Coronet} region, unless they are particularly
strongly obscured.

The remaining object is CrA~465, a brown
dwarf candidate with estimated spectral type M8.5.
It was not detected as an X-ray source by the automatic source
detection, but 
as will be discussed in \S \ref{bd.sec}, inspection of the
location of this object in the \textsl{Chandra} image
yields indications of a tentative
detection of very weak X-ray emission.


\subsubsection{IRS 3: a background giant rather than a YSO}

The bright infrared source IRS~3  (2MASS J19020491-3658564)
is usually considered to be a YSO associated with the CrA
star-forming region. 
In a spectroscopic and photometric study,
\citet{Nisini05} derived 
a spectral type K5--M0\,III, $L_\star = 0.3\,L_\odot$, $A_V = 10$~mag,
$M \sim 0.5\,M_\odot$, and an age $\sim 3$~Myr, but found no
no significant near-infrared excess and no signs of accretion.
Remarkably, IRS~3 remains undetected in the
\textsl{Chandra} data. 
We derive a 90\% confidence count rate upper limit of
$< 0.0158$~counts/ksec, that 
corresponds to an upper limit of
$L_{\rm X} < 3.2 \times 10^{27}$~erg/sec and  
 $\log\left(L_{\rm X}/L_{\rm bol}\right) < -5.6$.
This value would be most unusually low if IRS\,3 was a YSO:
in the COUP data 
more than 99.5\% of all young late-type stars in the Orion Nebula Cluster have
fractional X-ray luminosities above this level.
This non-detection therefore suggests that IRS~3 is {\em not}
 a young star, and this argument 
is supported by several other pieces of evidence:
First, IRS~3 shows no infrared excess.
Second,
the age of $\sim 3$~Myr, derived under the assumption that
the object is located in the CrA star-forming region, is considerably larger
than the ages of all other \textsl{Coronet} objects observed by \citet{Nisini05}.
Third, IRS~3 is the only one of the objects studied by \citet{Nisini05} with 
luminosity class III;
all other have luminosity class V (as expected for young stars).
Taken together, these considerations strongly suggest that
IRS~3 is {\em not} a YSO in the \textsl{Coronet} cluster, but rather a background giant behind the dark cloud.

\subsubsection{YSO candidates from \textsl{Spitzer}-IRAC  photometry}

\textsl{Spitzer} IRAC colors were used to classify the infrared sources
as described in \cite{Megeath04}. 
While class~I or class~II objects are probably
YSOs, the class~III objects are a mixture of YSOs which have already
dispersed their (inner) disks and unrelated field stars.
We thus consider objects with class~I or class~II SEDs and no previous
identification as new YSO candidates.
As discussed in \cite{Megeath04},
several factors may lead to incorrect classifications. For example,
some of the objects identified as class~I may be in fact
strongly reddened class~II objects, and background objects such
as planetary nebulae, AGB stars, and galaxies may be misidentified
as class~I or class~II objects.
The detection of X-ray emission at levels typical for YSOs
allows a clear distinction between YSOs and background objects of
the above-mentioned kind. We detect 14 of the 17 objects with Spitzer class~II SEDs, and 9 of the 10 objects with Spitzer class (0)I SEDs.

\paragraph{\bf Class~I objects:}

\textsl{Spitzer} photometry reveals three objects in the \textsl{Chandra} field of view
with class~I SEDs that were not identified as YSOs before.
Two of them, 190148.46--365714.5 and 190155.61--365651.1
(both remained undetected in the 2MASS images),
can be identified with X-ray sources.
Their X-ray detections strongly support the YSO status of these 
objects. With 22 and 41 source counts, respectively, these X-ray
sources are too faint for detailed spectral fitting, but
the high median energies of their source photons  of
4.1~keV and 4.5~keV are fully consistent with the hard spectra
expected for embedded class~I objects.

The third new object with \textsl{Spitzer} class~I SED  is the
infrared source B185836.1-370131 discovered originally by
\citet{Wilking97}.
It is invisible in the 2MASS $J$- and $H$-band images but is seen as a
 very faint source in the 2MASS  $K$-band image. 
This object coincides with sub-mm source SMM~2 from \citet{Nutter05},
but remains undetected in the \textsl{Chandra} data.
If it is truly a protostellar member of the CrA star-forming region, its
non-detection in the \textsl{Chandra} data could be related to very strong 
extinction, as suggested by the non-detection in the 2MASS images.
Since we cannot estimate the extinction to this object, no upper limit
to the X-ray luminosity can be determined.
However, we can ask how much extinction would be required to keep
it undetected in the \textsl{Chandra} data, if one assumes that it has 
an X-ray luminosity of $\sim 3\times 10^{30}$~erg/sec
(the mean value for the X-ray detected \textsl{Coronet} class~I objects
IRS\,1, 5, and 2).
Using PIMMS and assuming a plasma temperature of $30~$MK, we find 
that 
a  hydrogen column density of $2.2 \times 10^{24}\,{\rm cm}^{-2}$, 
corresponding to an extinction
of $A_V\sim 1100$~mag, would be required 
for this object to escape detection in the \textsl{Chandra} data.
This extreme extinction may perhaps be caused by occultation
of a massive circumstellar disk seen exactly edge-on, or
B185836.1-370131 may be very  deeply embedded
in a dense circumstellar envelope, and thus is perhaps a class~0 
object. 
An alternative possibility would be that it is a background AGB star 
(which would not be a strong X-ray emitter); in that case, however, 
its location
just behind a sub-mm cloud core would be a quite curious coincidence.


\paragraph{\bf Class~II objects:}

Three objects with class~II \textsl{Spitzer} SEDs remain undetected in the \textsl{Chandra} data:
LS~CrA~I  and  B185831.1-370456, two brown dwarf candidates 
which will be discussed in \S \ref{bd.sec}, and 
2MASS~J19020682-3658411. The later object is a relatively faint
infrared source and shows a near-infrared excess. It may be
a very-low luminosity (and correspondingly very-low mass, perhaps
sub-stellar) member of the CrA star-forming region suffering from particularly 
strong extinction, or an unrelated background AGB star.


\section{X-ray properties of the YSOs}

\subsection{X-ray variability and spectroscopic analysis}
\label{sect.xspec}
Since the short-term variability (as seen in the light curves
of the individual observations) of the \textsl{Coronet} X-ray sources
is discussed in \cite{Forbrich06a} and \cite{Forbrich06b}, 
we focus here entirely on the long-term variability defined
by the temporal sequence of the \textsl{Chandra} observations,
covering a period of nearly five years.

\begin{figure*}
\includegraphics*[width=18cm, bb=60 250 590 700]{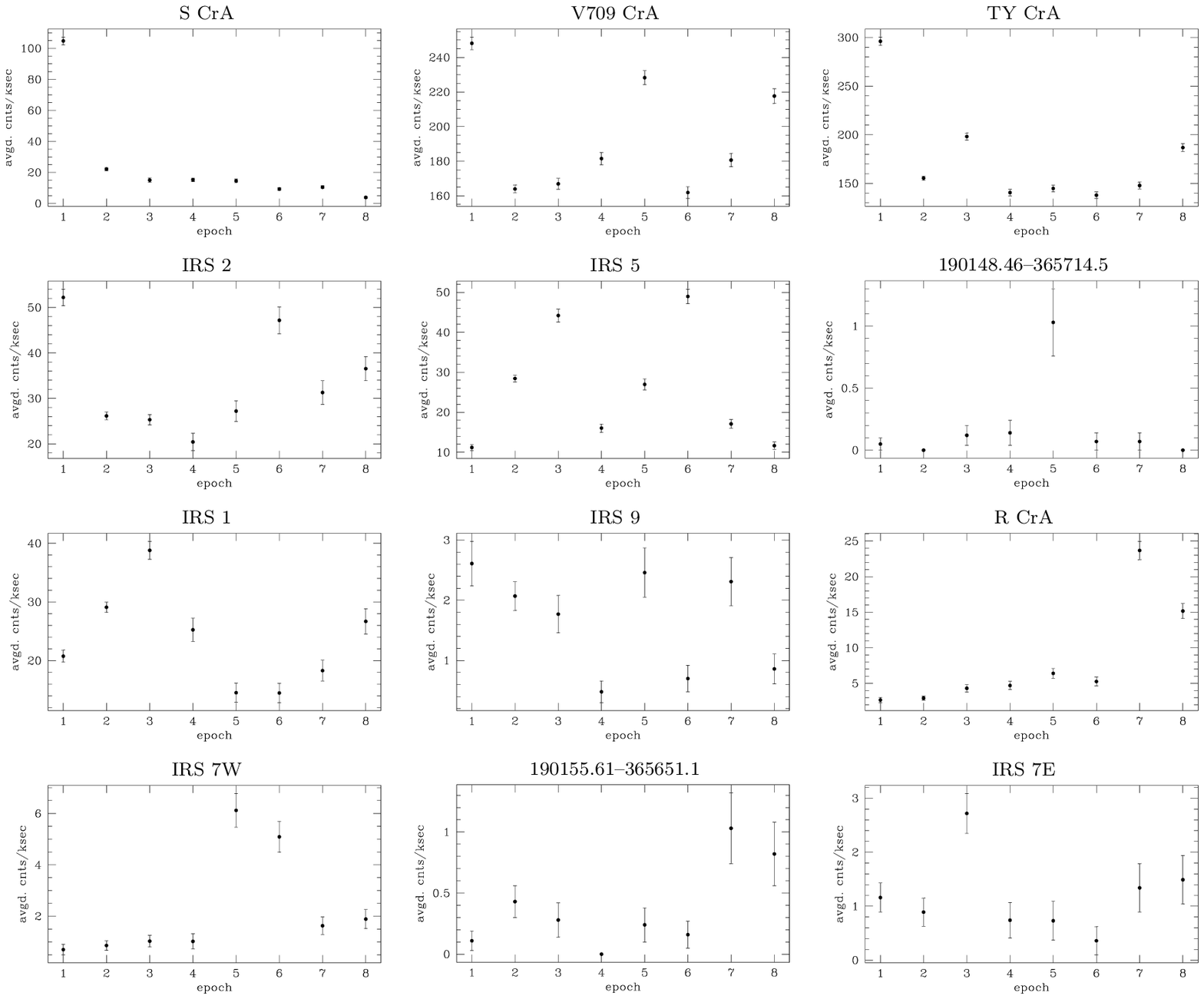}
\caption{Long-term evolution of the X-ray emission of selected sources
in the \textsl{Coronet} region derived from the \textsl{Chandra} observations obtained
in October 2001 (epoch 1), June 2003 (epoch 2), June 2004 (epoch 3),
 and August 2005 (epochs 4 to 8).
The dots show the mean count rates during each of the individual
observations, which have exposure times between  15 and 40 ksec. The count rates were determined from counts between 0.2~keV and 8~keV and were corrected for effective exposure times using exposure maps.
}
\label{lightcurves.fig}
\end{figure*}

For the present analysis we determined for each source the
mean count rates during each of the 8 individual \textsl{Chandra}
observations.
While many sources show only small and often statistically 
insignificant variations, strong variability is seen in 
some of the YSOs.
The more interesting long-term lightcurves are shown in 
Fig.~\ref{lightcurves.fig}, and the
variability of individual objects will be
discussed below.

\bigskip

A detailed analysis of the X-ray spectra of all sources with 
optical/infrared counterparts
was performed with the \textsl{Sherpa} package in CIAO.
The spectra were fitted with one- and two-temperature 
optically thin thermal plasma models plus an intervening
absorption term. 
We used the XSPEC models ``apec'', 
assuming a uniform density plasma with 0.3
times solar elemental abundances,
and ``wabs'' for the absorption model.
Spectral fits were carried out ignoring energy bins outside an energy range of 0.2--10~keV. As X-ray spectral fits sometimes suffer from ambiguities
in the spectral parameters,
special emphasis was placed on a careful scanning of the
parameter space in order to find the best fit model.
For this, we employed the \textsl{monte-lm} algorithm
implemented in \textsl{Sherpa}, which performs hundreds of fitting
runs per spectrum, each one using a different set of
randomly chosen starting values for the fitting parameters.

\begin{figure*}
\includegraphics[width=18.0cm, bb=60 250 590 700]{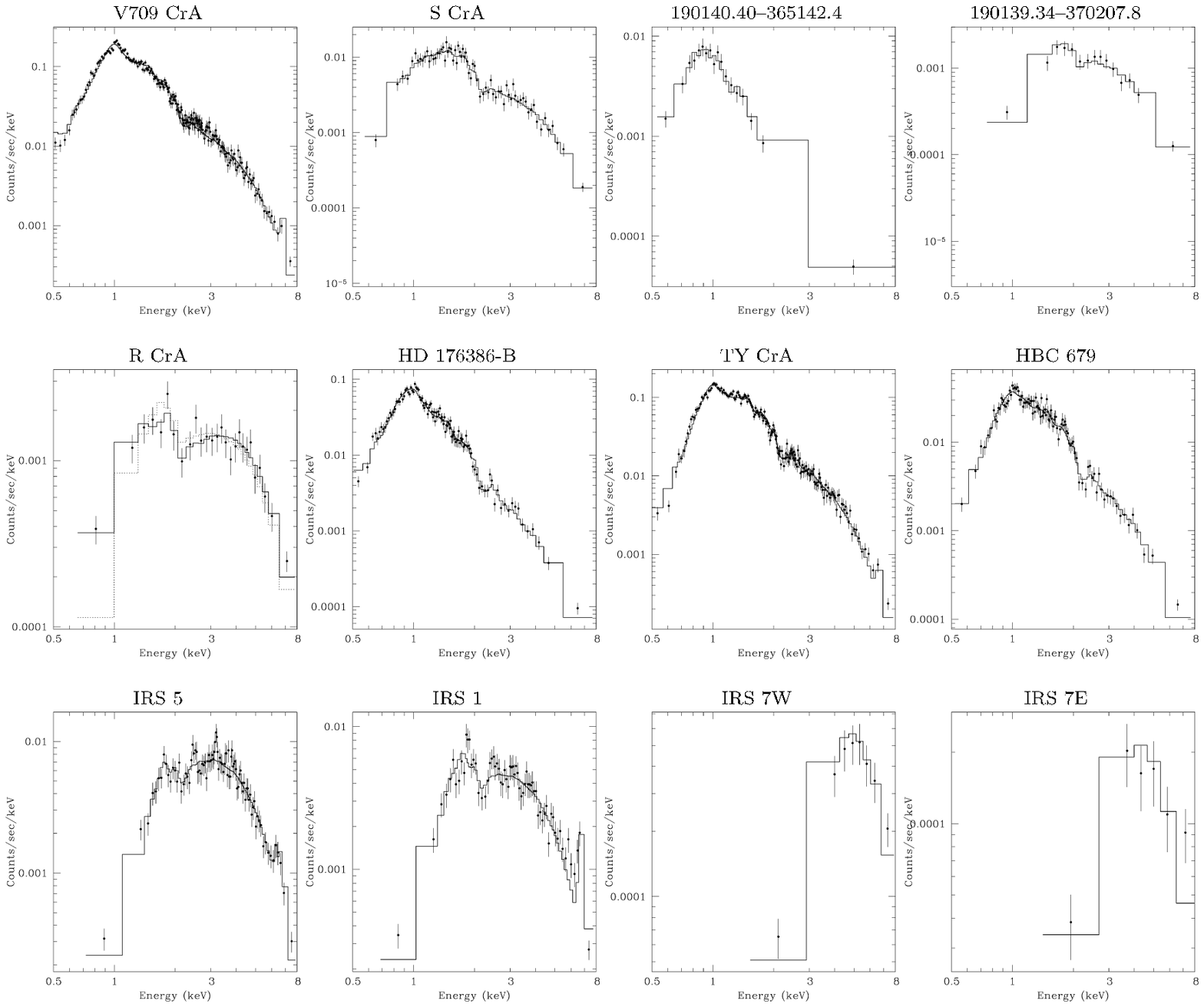}
\caption{
\textsl{Chandra} X-ray spectra of selected
YSOs in the \textsl{Coronet} region. 
The solid dots with error bars show the observed spectra,
while the histogram lines show the best fit models.
The first row shows two pairs of class III / II objects:
V~709 (class~III) \& S~CrA (class~II), and 190140.40 (class~III)
\& 190139.34 (class~II), which illustrate the systematically harder
spectra of the class~II objects.
The first plot in the second row shows the very hard spectrum of the 
Herbig Ae star R CrA; the dotted line shows the one-temperature fit model,
the solid line corresponds to the ``two temperature -- two absorption'' model
(for details see text).
 The next three panels compare the spectrum of the companion to the B star
HD~176386 and the spectrum of the TY~CrA multiple system to the spectrum
of the class~III T Tauri star HBC~687; the  
similarity of these spectra suggest that the true sources of X-ray emission
apparently observed from these B stars are most likely young late-type stellar
companions.
The third row shows three class I objects and
finally the class 0/I protostar IRS~7E. }
\label{spectra.fig}
\end{figure*}

Spectra of sources with less than 1000 counts  
were generally well fitted with a single-temperature plasma 
model, 
for stronger sources and sources for which the single-temperature
model did not provide an acceptable fit,
a two-temperature model was employed.
The spectral analysis yielded plasma temperatures and hydrogen 
column densities
and was also used to compute the 
intrinsic (extinction-corrected) X-ray luminosity by
integrating the model source flux over the $0.2\!-\!8$~keV band.
The results are listed in Table~\ref{table_fit}.
Some representative examples of spectra are shown in 
Fig.~\ref{spectra.fig}.
For X-ray sources  with less than $\sim 50$ counts, the spectral
fitting procedure often does not allow to reliably determine
the spectral parameters.
In these cases,  the incident flux
 at the telescope aperture as determined by {\rm{\it acis\_extract}}
(see Table \ref{acis_extract.tab}) provides at least 
a rough estimate of the {\em observed} 
(i.e.~{\em not} extinction-corrected) source luminosity.

\subsection{X-ray properties of different object classes}

\subsubsection{B- to F-type stars}

\paragraph{\bf TY CrA:}\,\,\,
The B8e star TY~CrA is (at least) a quadruple system:
in addition to the two spectroscopic companions with separations
of 0.07~AU and 1.2~AU and estimated spectral types
of $\sim$K2 and late F to early K \citep{Casey95,Corporon96,Casey98},
a visual companion at 
a projected separation of 41~AU ($0.3''$) and of spectral type
$\sim$~M4 was found by
\citet{Chauvin03}.
With 21026 counts, the TY~CrA system represents the second-brightest 
X-ray source in the field of view. 
Since the individual components 
cannot be resolved in the X-ray data, their contributions 
to the observed X-ray emission cannot be discerned.
However, it is interesting to note that
the observed X-ray properties agree very well with the
expected X-ray emission of the three late-type companions,
assuming that these stars have X-ray characteristics similar
to those of other stars with similar age and mass:
the median X-ray luminosities of young G- and early K-type stars
in the COUP sample are
$2.5\times 10^{30}$~erg/sec, and for young M-type stars
$3.0\times 10^{29}$~erg/sec 
\citep{Preibisch_coup_evol}. Thus, the {\em expected}
 combined X-ray luminosity
of the three companions of TY~CrA is
$5.3\times 10^{30}$~erg/sec, a value that is 
very close to the observed system
X-ray luminosity of $5.0 \times 10^{30}\,{\rm erg/sec}$.
Also, the plasma temperatures of $T_1 = 10$~MK and $T_2 = 27$~MK
derived from the fit to the X-ray spectrum of the TY~CrA system
are in the prototypical range observed for G- and early K-type T Tauri
stars \citep[see, e.g.,][]{Preibisch_coup_orig}.
The observed X-ray emission from the TY~CrA system can thus
be very easily understood as
originating from the late-type companions of TY~CrA;
the data provide no direct hint towards possible X-ray emission
from the B8e star. Of course, we cannot rule out the
possibility that (some fraction of) the observed X-ray emission
may nevertheless come from the Be star, but 
the data provide no indications for this assumption.


\paragraph{\bf HD 176386:}\,\,\,
This B9\,IVe star 
has a visual companion
at $\sim 4''$ separation \citep{Jeffers63}.
The strong decrease of the brightness ratio between primary and 
companion from optical to near-infrared wavelengths of
($\Delta [V, J, H, K] = [6.1, 0.76, 0.68, 0.45]$~mag
according to \citet{Turon93} and the 2MASS point-source catalog) 
suggests that the companion is of substantially later spectral type
and thus most likely
a low-mass ($M \la 2\,M_\odot$) star.

A strong X-ray source with 7720 counts is 
perfectly centered on the position of the companion to HD~176386, 
whereas there is no evidence for emission from the B star
in our data.
The spectral fit yields parameters
($T_1 = 9$~MK, $T_2 = 21$~MK,
$L_{\rm X} = 1.1 \times 10^{30}\,{\rm erg/sec}$) which are
very typical for K- and early M-type T Tauri stars.

In order to derive an upper limit to the possible X-ray emission from
the primary B-star HD~176386, we extracted counts in a $1''$ radius 
aperture centered
on its optical position.
There are 21 photons in this region, which however is strongly affected 
by the wings of the PSF of the X-ray emission from the companion.
Comparison with nearby ``background regions'' at the same radial distance
from the strong X-ray source yielded an expected background of
22 counts in a $1''$ radius aperture. The 90\% confidence upper limit
to the number of counts from HD~176386 is 8.4 source counts.
After correction for the small extraction region, which would contain
only about 30\% of the flux from a point source at the 
off-axis angle of $\sim 6'$, we derive an upper limit of $<0.21$~counts/ksec
for the source count rate; with $A_V = 1$~mag and $L_{\rm bol} = 58\,L_\odot$
\citep{Bibo92} we derive upper limits of 
$L_{\rm X} < 4.5 \times 10^{27}$~erg/sec and 
$\log\left(L_{\rm X}/L_{\rm bol}\right) < -7.7$.


\paragraph{\bf  R CrA:}\,\,\,
The Herbig Ae star R CrA 
shows an extremely strong infrared excess.
According to the analysis of \citet{Acke04},
the stellar luminosity, derived by fitting and integrating a model 
for the stellar photosphere to the de-reddened photometry,
is only a very small fraction of the bolometric luminosity, i.e.~the
total integrated luminosity of the SED.
The fact that the derived stellar luminosity would place R~CrA 
{\em below} the
main-sequence in the HR-diagram, shows that the stellar parameters
are very uncertain. Perhaps, the central star is deeply embedded 
in circumstellar material and seen only in
scattered light. In that case, the stellar luminosity and
also the derived extinction of $A_V = 1.33$~mag would be considerably 
underestimated.

The X-ray emission from R~CrA is strongly variable and has risen
to considerably higher levels in the last two epochs of our monitoring
\textsl{Chandra} observations (Fig.~\ref{lightcurves.fig}, see
also X-ray lightcurves in \citealp{Forbrich06a}).
The lightcurves of the individual August 2005 \textsl{Chandra} observations 
(shown in \citealp{Forbrich06b}) reveal numerous flare-like peaks, suggesting
that the source is more or less continuously flaring.
Our spectrum of R~CrA contains 981 source counts.
The extreme hardness of the spectrum, providing clear evidence for 
the presence of very hot plasma,
was already noted in \cite{Forbrich06a} and is clearly confirmed.
A one-temperature model cannot produce an acceptable fit to the
observed spectrum; while the hard part ($> 2$~keV) of the spectrum 
is rather well fitted, a clear excess of counts remains at 
and just below 1~keV (this model is shown by the dashed line histogram
in the spectrum in Fig.~\ref{spectra.fig}).
A two-temperature model provides a formally acceptable fit
($\chi^2_r = 0.83$), but 
the resulting spectral parameters are very dubious:
in addition to an extremely hot plasma component, for which
only a lower bound to the temperature ($T_{2} > 660$~MK)
can be established, the fit
yields an extremely strong and cool low-temperature component of 
$1.3$~MK and strong extinction of 
$N_{\rm H} = 2.3 \times 10^{22}\,{\rm cm}^2$.
The emission measure of this alleged low-temperature plasma 
component would be more than 2000 times larger than that of the 
high-temperature component, and its  (extinction-corrected)
X-ray luminosity
would be $1.6\times 10^{32}$~erg/sec, orders of magnitude 
higher than the luminosity of the
high-temperature component of 
$4.3\times 10^{29}$~erg/sec.
This extremely high X-ray luminosity makes it very unlikely
that this very soft spectral fit component represents
soft X-ray emission due to accretion shock emission.
Rather, we strongly suspect
that this fit result is an example of the
highly nonlinear interaction of very low-temperature plasma
components and strong extinction
\citep[see, e.g., the discussion in][]{Getman05a}.
For high hydrogen column densities, such a very low-temperature 
plasma component is almost entirely absorbed, and thus 
any uncertainties in the lowest energy bins
of the spectrum can lead to large overestimates of the
extinction-corrected X-ray luminosity. 

In an attempt to find a spectral fit solution
that avoids this kind of inference of a very luminous,
but heavily absorbed, ultrasoft component, we considered a 
spectral model in which both of the 
two plasma components have individual extinction factors
rather than a common extinction factor for both components
(model = wabs$_1 \times$ apec$_1 +$ wabs$_2 \times$ apec$_2$,
instead of the ``usual'' model = wabs $\times$ [apec$_1 +$apec$_2$]).
This model provides a very good fit ($\chi^2_r = 0.66$)
for $N_{\rm H, 1}=\left(1.5 \pm 0.1\right)\times 10^{22}\,{\rm cm}^{-2}$
($A_V = 7.25$~mag), $T_1 = \left(9.3 \pm 0.8\right)$~MK, and
$N_{\rm H, 2}=\left(4.0 \pm 0.3\right) \times 10^{22}\,{\rm cm}^{-2}$    
($A_V = 20$~mag), $T_2 > 607$~MK, and 
is shown by the solid line histogram
in the spectrum in Fig.~\ref{spectra.fig}.
The extinction-corrected
X-ray luminosities for the two spectral components
are  $L_{\rm X, 1} = 2.4 \times 10^{29}$~erg/sec
and $L_{\rm X, 2} = 5.3 \times 10^{29}$~erg/sec.

The requirement of different extinction values for the two 
temperature components in the spectrum of R~CrA is
similar to the case of FU~Ori,  
for which \citet{Skinner06} found that the hot plasma component also
requires a considerably larger hydrogen column density than
the low-temperature component to fit the observed spectrum.
They argued that the hot
component represents coronal emission that is strongly absorbed
in accretion streams or a strong stellar wind.
Why the cooler spectral component is less absorbed,
remains unclear; perhaps it originates
from a different location. 
An alternative explanation would be that the X-ray emission comes
from two different, close and thus unresolved objects
(perhaps late type companions to R~CrA) for which the
extinction along the line-of-sight is different.
Although these explanations remain quite speculative, it is 
interesting that R~CrA and FU~Ori are both very strongly accreting,
and their peculiar X-ray spectra may therefore be in some way
affected by accretion processes.
If the X-ray emission originates from R~CrA (and not from an
unresolved companion)
the extinction derived from the X-ray spectrum
($A_V \ge 7$~mag) supports the
suspicion that the optical extinction value of $A_V = 1.33$~mag 
is a serious underestimate.


\paragraph{\bf T CrA:}\,\,
This F0e star is particularly interesting 
because its spectral type is very close to
the upper boundary for stars with convective envelopes.
For the stellar parameters of T~CrA  as listed in
\citet{Acke04}, the models of \cite{Siess00} suggest an extremely
shallow convective envelope with relative thickness
of $\Delta R_{\rm env}/R_\star =
0.8\%$. An interesting question is whether a magnetic dynamo may
work in such an extremely thin convection zone.

From an analysis of the first \textsl{Chandra} observation
of the \textsl{Coronet} (ObsID=19),
\citet{Skinner04} reported a tentative X-ray detection of T~CrA 
with $4\pm2$ counts in these data.
In our  much deeper dataset we clearly detect X-ray emission
from T~CrA, although again only as a rather faint source (14 source counts).
A fit of the X-ray spectrum,
with the hydrogen column density fixed at the
value corresponding to the extinction of $A_V = 2.45$~mag
as derived by \citet{Acke04}, suggests a plasma temperature
of $\sim 11$~MK and yields an X-ray luminosity of
$\sim 5 \times 10^{27}$~erg/sec.
This F0 star is thus clearly a much weaker X-ray emitter than
the G type T Tauri stars in the \textsl{Coronet} and in other
young clusters \citep[see, e.g.,][]{Preibisch_coup_orig}.
This suggests that T~CrA is the hottest object in the \textsl{Coronet} region with coronal X-ray activity driven by a dynamo in a very shallow convection zone.


\subsubsection{T Tauri Stars}

\paragraph{\bf S CrA :}\, 
This classical T~Tauri star of spectral type K3 has a
faint companion ($\Delta K = 3.27$), 
also a classical T~Tauri star,
 with spectral type M0,  located at a separation
of $1.3''$ from S~CrA \citep{Prato03, McCabe06}.
The \textsl{Chandra} data yield 3020 source counts, and their spatial
distribution
is consistent with a single source
at the position of the primary; there is no elongation
along the direction toward the companion (along
position angle $149\degr$).
The spectral parameters are prototypical for T Tauri stars.

\paragraph{\bf HBC 679 :\,}
This weak-line T Tauri star of spectral type K5
has a companion of spectral type  M3 at a separation of $4.5"$
\citep{Prato03}.
The \textsl{Chandra} source with 5468 counts is centered at the
position of the primary,
  but there is an elongation in the direction of the secondary.
  Since the source is located close to the edge of the field of view
of the \textsl{Chandra} image, where the PSF is already degraded,
the contributions of the individual components cannot be
reliably disentangled.



\subsubsection{The class 0 protostar candidate IRS 7E}

The infrared source IRS~7E has many properties that are characteristic
for class~0 protostars. Its SED is dominated by the strong submillimeter emission \citep{Nutter05}, and possibly the source has a disk and an outflow \citep{Anderson97,Groppi04}, although a definitive attribution is currently difficult due to the high source density in the region and the comparably low angular resolution. \citet{Harju01} find evidence for a radio jet emanating from IRS~7E.
IRS~7E is detected in all four IRAC bands of the Spitzer images.
Based on this and new high-angular resolution submillimeter data, \citet{Groppi07} conclude that IRS~7E (their source SMA~1) is a transitional class~0/I source, thus the youngest of the sources discussed here.
This result supports the notion that IRS~7E is different (i.e.~in an earlier
evolutionary state) from the class~I objects.

The X-ray detection of IRS~7E (originally reported by \citealp{ham05b})
represents up to now the only reliable case for high-energy emission
from this early evolutionary stage.
The long-term lightcurve (see Fig.~\ref{lightcurves.fig})
suggests considerable variability, but (probably due to the
low count rate) no individual flares are detected in the
individual \textsl{Chandra} observations.
The X-ray spectrum of IRS~7E 
(see Fig.~\ref{spectra.fig}) is very hard; the spectral fit
suggests an extinction of $A_V \sim 74$~mag,
yields a plasma temperature of $80$~MK, and gives
an extinction-corrected X-ray luminosity
of $\sim 4 \times 10^{29}$~erg/sec.
These parameters are roughly consistent with those derived
by \citet{ham05b} from their XMM data for the phase of 
``constant'' emission before the flare.
The very high plasma temperature clearly shows that the X-ray 
emission is dominated by magnetic processes, suggesting that
magnetic activity starts already in extremely early stages
of (proto)stellar evolution.

\subsubsection{Class I protostars}

\begin{figure}
\begin{center}\includegraphics[width=7cm, bb=20 150 550 670]{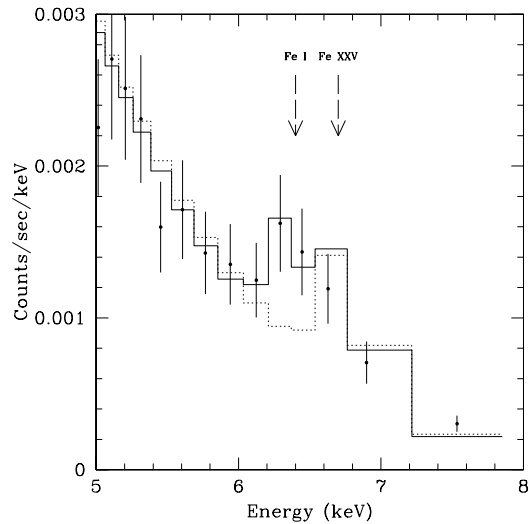}\end{center}
\caption{Detail of the X-ray spectrum of IRS 5 around the $6.7$~keV iron line,
showing the strong fluorescent emission at 6.4~keV. The dashed line shows the best-fit model without taking into account a line at 6.4~keV.}
\label{irs5.fig.specdet}
\end{figure}

X-ray emission from seven class~I protostars is detected
in our dataset: IRS~1, 2, 5, 7\,W, 9, 190148.46--365714.5,
and 190155.61--365651.1. The X-ray properties of these objects
are similar to those of other X-ray detected class~I protostars.

The class~I object IRS~5 is especially interesting.
The object is a close binary with a separation
of $0.9''$ \citep{Chen93,Nisini05}. The X-ray emission is 
marginally resolved in the Chandra data,
with the infrared brighter component being also the stronger
X-ray source. Due to the 
PSF overlap, the individual components cannot be
reliably resolved and thus
we study here only the composite spectra and lightcurves.
The X-ray spectrum of IRS~5 (Fig.~\ref{spectra.fig}) shows a prominent
emission line at
6.7~keV, which is the K-line from highly ionized iron (Fe XXV).
A closer look at the spectrum (Fig.~\ref{irs5.fig.specdet})
reveals significant excess emission around 6.4~keV, which
we identify as 6.4~keV line emission
from neutral to low-ionized iron.
The addition of a 6.4~keV emission line to the spectral model
yielded a good fit for a line width of $44\pm15$~eV.
The quality of the spectral fit with this 6.4~keV line
(reduced $\chi2$ of 0.91 for 84 DOF) is clearly better
than without this line  (reduced $\chi2$ of 0.97 for 87 DOF);
an $F$-test shows that the null hypothesis, i.e.~the assumption
that the observed 6.4~keV excess is due to random noise,
can be rejected with 99\% certainty.
Considering only the $5- 8$~keV range
of the spectrum, as shown in Fig.~\ref{irs5.fig.specdet}, the
reduced $\chi2$ changes from 1.45 (10 DOF) for the fit without a 6.4~keV
line to 0.78 (7 DOF) for the fit with a 6.4~keV line.

Similar 6.4~keV iron fluorescence lines have been observed in some other
YSOs \citep[e.g.,][]{Imanishi01a,Favata05,Tsujimoto05}.
The presence of the 6.4~keV K-line from cool iron
in addition to the 6.7~keV K-line from hot iron
can be explained as
fluorescence emission from cool material that
is irradiated by the hard X-ray emission from the YSO.
However, it is not immediately clear {\em where} this cool,
fluorescing material is located.
One obvious possibility would be fluorescence from irradiated
material in a circumstellar disk, but the emission may in
principle also originate from interstellar material somewhere
along the line of sight.
As discussed in \citet{Tsujimoto05}, the width of the
6.4~keV line can help to distinguish between these different
possibilities (see also \citealp{Favata05}). For fluorescence emission from
interstellar
material along the line of sight, one would expect a line width
that should be considerably smaller than 10~eV (formula~(4) in
\citealp{Tsujimoto05}).
The observed line width of 44~eV therefore clearly suggests
that the fluorescent emission comes from dense local, i.e.~circumstellar
material of higher column
density, most likely from the irradiated circumstellar disk of IRS~5.
The strong X-ray irradiation of circumstellar disk
material has important consequences for the
physical processes in the circumstellar dust and gas
and the evolution of proto-planets
\citep[e.g.,][]{Glassgold05,Wolk05}.

\subsubsection{Brown Dwarfs\label{bd.sec}}

\begin{table*}
\centering
\caption{Results for the Brown Dwarf candidates (see Section \ref{bd.sec}).
}
\label{bd_tab}       
\begin{tabular}{rrccrrrrrr}
\hline\noalign{\smallskip}
 Name &Ref&  SpT &  $\log\left(\frac{\displaystyle L_{\rm bol}}{\displaystyle L_\odot}\right)$ &  M &  $A_V$ & source & $T_{\rm X}$ & $L_{\rm X} $\,\,\,\,\,\,\,\, &  $\log\left(\frac{\displaystyle L_{\rm X}}{\displaystyle L_{\rm bol}}\right)$ \\
      &  &    &  & $[M_\odot]$ & [mag] & counts  & [MK] & [erg/sec] & \\
\noalign{\smallskip}\hline\noalign{\smallskip}
B185815.3-370435 &1&   &  &    & $\sim 7$ &   $ <3.2 $ &  & $ < 2.6\times 10^{27}   $ & \\
B185831.1-370456 &1,2& M8.5 & $-2.6 $ & $\sim 0.025$  & $\sim 0 $ & 9& &   $\sim8.0\times 10^{26}$ & $\sim  -4.1 $ \\
B185839.6-365823  &1&   & $-3.1 $  & &  $\sim 13$   &  47& $> 350 $ & $\sim  2.7 \times 10^{28}$ & $\sim -2.1 $\\ 
%
B185840.4-370433  &1&   & $-2.2 $ &   &  $\sim 12$&  6  &  & $\sim 1\times 10^{28}$& $\sim -3.4$\\
B185853.3-370328 &1&   &$-2.2 $ &   &  $\sim 12$& $<2.1$ & & $ <4.5\times 10^{27}$ & $<-3.7$\\
LS-RCrA 1 &3& M6.5-7 & $-2.6$ & $\sim 0.04\!-\!0.08$ & $\sim 0.5$ & $<10$ &  &  $ <1.1\times 10^{27}$ & $< -3.9 $\\
LS-RCrA 2   &3& M6   & $-1.5$  & $\sim 0.08$ &  $\sim 1.2 $ &   113 & 4.5 &  $ 3.1 \times 10^{28}$ & $-3.6$ \\  
%
ISO CrA 143 &4& $\approx$ M8: & $-2.0  $ & $\sim 0.025$ & $\sim 2.3$ & 88 &  8 & $2.9 \times 10^{28}$ & $-3.2$ \\
\noalign{\smallskip}\hline
\end{tabular}\smallskip

References: 1:~\citet{Neuhaeuser99}; 2:~\citet{Lopez05};
3:~\citet{Fernandez01}; 4:~\cite{Olofsson99}
\end{table*}

X-ray emission has been detected from numerous {\em young} BDs
\citep[e.g.,][]{Neuhaeuser98,Imanishi01b,PZ01,PZ02,Tsuboi03,Preibisch_coup_bd},
but the origin of their activity is still not well understood.
The CrA star-forming region contains a number of very low-mass objects,
some of which are most likely young brown dwarfs (BDs).
Several objects have been identified as BD candidates
in different studies \citep{Wilking97,Neuhaeuser99,Olofsson99,Fernandez01,Lopez05},
but, unfortunately, in no case a fully 
reliable spectroscopic classification as BD is available,
because either the spectral types were estimated from
(narrow-band) photometry and are thus quite uncertain, or the 
objects are very close to the stellar/sub-stellar boundary.
Therefore, all objects discussed here are BD candidates, not
bona-fide BDs.

There are eight
BD candidates in the field of view  of the \textsl{Chandra} image.
Three of them (B185839.6-365823, LS-RCrA 2, and ISO CrA 143) 
are among the X-ray sources detected  by \textsl{wavdetect}. 
For the remaining five BD candidates we have performed a detailed 
investigation of
the corresponding positions in the \textsl{Chandra} image. For each object we defined
a source region centered at its optical position with a radius of
$3''$ and a corresponding background region 
 as annulus with inner radius $5''$ and outer radius $10''$.
Then we determined the numbers of 
detected counts in the source and background regions and compared
the number of background counts scaled by the corresponding area
to the number of counts detected in the source regions.
For two objects (B185840.4-370433 and B185831.1-370456)
the number of counts in the source region exceeded the number of expected
background counts with at least 90\% confidence, tentatively indicating
the presence of very weak X-ray emission.
For the remaining 3 objects, upper limits to their count rates 
and (if information on extinction was available)
also to their X-ray luminosities
were determined as described above.
The results of this analysis are listed in Table~\ref{bd_tab}.

The X-ray luminosities and 
fractional X-ray luminosities of the
young BD candidates in the CrA star-forming region  are similar to those of the 
low-mass stars, 
and thus there is no evidence for changes in the magnetic activity 
around 
the stellar/substellar boundary. 
In two of the three objects which yielded enough counts for spectral
analysis, the derived plasma temperatures are 
in the lower range of plasma temperature found for young stars,
consistent with previous findings.
On the other hand, the BD candidate B185839.6-365823 shows
a rather hard spectrum and the fit suggests a very high 
plasma temperature. Although the S/N of the spectrum is quite 
low and thus no reliable determination of the spectral parameters
is possible, the median photon energy of 3.4~keV already
suggests a relatively hard spectrum.
We also note that this objects shows strong variability in the
long-term lightcurve, so its hard spectrum may be related
to X-ray flaring.


\subsection{Comparison of the X-ray properties of different
object classes}

\begin{figure}
\begin{center}\includegraphics[width=8.5cm]{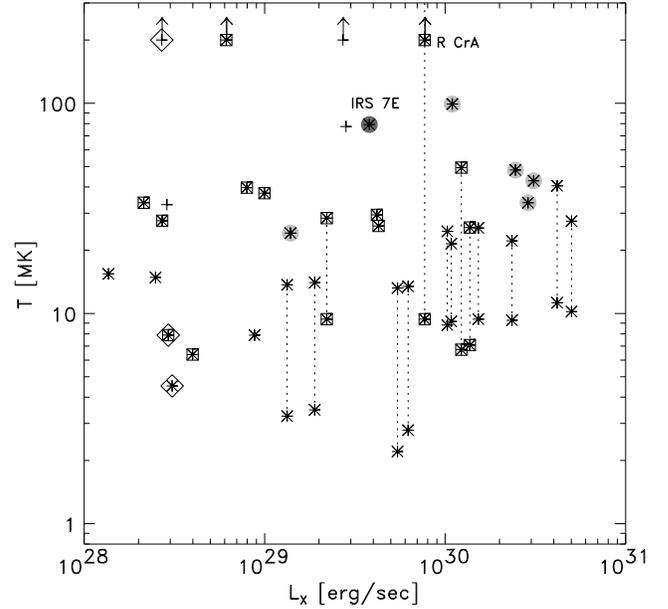}\end{center}
\caption{Plasma temperatures versus X-ray luminosities for the
CrA members with sufficient counts for X-ray spectral fitting.
The different SED classes are shown with different symbols:
the class~0/I protostar IRS~7E as a dark dot,
class~I objects as grey dots,
class~II objects as open boxes,
class~III objects as asterisks, and objects without SED
classification as crosses.
Brown dwarf candidates are surrounded by diamond symbols.
 For objects with 2T fits, the
high and low-temperature values are connected by the dashed lines.
} \label{lx_tx.fig}%
\end{figure}

In Fig.~\ref{lx_tx.fig} we plot the plasma temperatures
versus X-ray luminosities for the YSOs in the Coronet cluster.
We first consider the derived plasma temperatures, which yield
information about the X-ray emission process in the sense
that hot ($\ga 10$~MK) temperatures require magnetic processes
while the expected signature of emission from accretion shocks would
be at much cooler ($\la 3$~MK) temperatures.
Only two objects show cool plasma
components below 3~MK; both are class~III objects, i.e.~not
accreting. For all potentially accreting objects (IR classes 0,
I, or II) the lowest plasma temperatures are above 6~MK.
We thus conclude that the derived plasma temperatures provide
no hint to X-ray emission from accretion shocks.

Next, we compare the plasma temperatures
derived for objects in different evolutionary stages.
While the source numbers are too low for statistically sound conclusions,
the plot reveals that
Class~II objects tend to show systematically hotter plasma temperatures
than class III objects. This effect is also illustrated in the
comparison of the spectra for V709 CrA (class~III) to 
S~CrA (class~II) and 190139.34 (class~III) to 190140.40 (class~II) 
in Fig.~\ref{spectra.fig}. 
We note that a similar spectral difference was found by
\citet{Flaccomio06} for the YSOs in the NGC~2264 star-forming region.

Another interesting aspect is that the class~I objects in turn tend to
show systematically higher plasma temperatures than class~II objects.
This is especially interesting in the context of the recent debate whether
class~I and class~II objects are truly in different evolutionary
stages,
or whether the classification is affected by other factors such as
the inclination under which the YSO is seen
\citep[see discussion in][]{White04,Eisner05,Doppmann05}.
In addition to the difference in plasma temperatures of
class~I and II objects, we also note that our long-term lightcurves 
(see Fig.~\ref{lightcurves.fig}) suggest that 
the class~I objects display
stronger levels of variability than class~II objects.
Similar differences in the plasma temperatures and
levels of variability between class~I and class~II/III
objects were found by Imanishi et al.~(2001a,b) for the
 $\rho$ Ophiuchi dark cloud.
These differences in the characteristics of the X-ray emission
 support the notion
that class~I and class~II objects are truly different.


\section{Diffuse X-ray emission} 

\begin{figure}
\begin{center}\includegraphics[width=8.5cm]{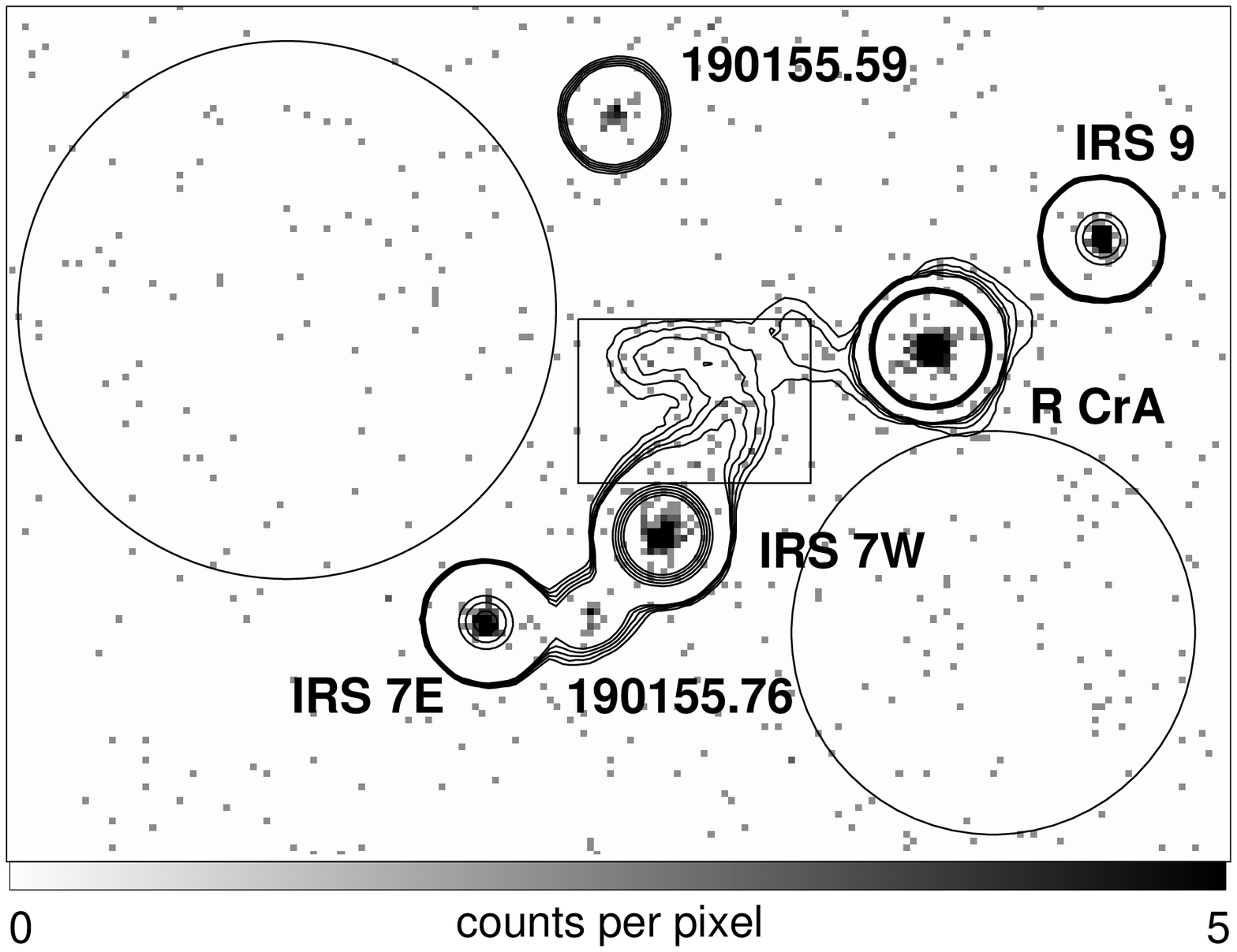}
\end{center}
\caption{\textsl{Chandra} X-ray image 
of the central region ($89'' \times 62''$ field of view), showing 
the
diffuse emission between IRS~7W and R~CrA. The greyscale shows
the $3-6.5$~keV band image with the original $0.49''$ pixels.
The contours
show the same image smoothed with a 10 pixel FWHM Gauss filter.
Contours are drawn at levels of 0.07, 0.08, 0.09, 0.1, 0.11
(to trace the diffuse emission) and 0.6, 0.7, 0.8, 0.9, 1.0
(to outline point sources.) 
The box indicates the source region used to characterize the diffuse emission while the two circles show the background regions.
} \label{diffuse_fig}%
\end{figure}

Inspection of the \textsl{Chandra} image reveals an excess of counts
in a region
north of the class~I object IRS~7\,W and east of R~CrA
(see Fig.\ref{diffuse_fig}). The box-shaped region contains 102 
counts, whereas the local background,
estimated from two nearby source-free regions, should contribute
only $\sim 38$ counts. The excess of 64 counts apparently represents 
diffuse X-ray emission.
Comparing the distribution of photon energies within this region 
to that 
in the background regions reveals a statistically significant
excess in the $3-6.5$~keV band.

The available optical and infrared images show no objects
or other structures that may be related to this emission.
We note that 
the continuum-subtracted [S II] image presented by \citet{Wang04}
shows a small arc of emission that seems to roughly coincide with
the diffuse X-ray emission. However, as there is much stronger and 
extended [S II] emission to the east of this patch, it is not
clear whether the X-ray emission is actually related to this
[S II]  emission. Furthermore, if the X-ray emission were related 
to the 
jets and outflows in this region, one would expect a quite soft
X-ray spectrum, with most of the detected photons at energies
of $\la 1$~keV and not $\ga 3$~keV as observed.

Another possibility would be that we see (perhaps scattered)
X-rays originating
from one or several embedded sources in this area. 
The source of the emission
could be IRS~7W, R~CrA or perhaps other, extremely deeply embedded
protostars. In this context, it is interesting to note that
\citet{Choi04} found four 6.9~mm sources near IRS~7W, which
may be very deeply embedded YSOs (none of them has a counterpart in 
any of the Spitzer images). Although none of these radio sources
coincides with the region of extended X-ray emission,
there may be further, still undetected, deeply embedded YSOs in 
this region, which could be the source of the X-ray emission.




\section{Conclusions and Summary}

The main results
of our very deep X-ray study of the \textsl{Coronet} cluster
can be summarized as follows:

The observed X-ray properties of the YSOs in the \textsl{Coronet} region
are fully consistent with coronal magnetic activity.
We find no indications for X-ray emission from
accretion- or jet-shocks: in our X-ray spectral analysis, we
find neither significant plasma components at temperatures below
$\sim 3$~MK, nor indications for soft ($\la 1$~keV) excesses for any of the
accreting stars.
This is consistent with results from other X-ray studies
\citep[e.g.,][]{Preibisch_coup_orig,Stassun07} showing 
that the bulk of the observed X-ray emission in most T Tauri stars
cannot originate in accretion shocks.
However, it has clearly to be noted that this result
does not exclude the possibility that accretion shocks
may produce some fraction of the X-ray emission in some of the
Coronet stars. A relatively weak soft excess in the X-ray spectrum of
a star suffering from more than a few magnitudes of visual extinction
would hardly be detectable in the data.

The observed tendency that the class~I objects exhibit a higher
degree of X-ray variability than the older class~II and III objects
may be a consequence of magnetic reconnection events in
the proto(star)-disk magnetic fields, causing frequent strong
flares \citep[see, e.g.,][]{Montmerle00}.
The apparent differences in the X-ray properties
of class~I versus class~II/III objects supports the assumptions
that class~I objects are truly in an earlier evolutionary stage.
Finally, the
high plasma temperatures of the class~0 and class~I protostars
clearly show that the X-ray emission of these extremely young objects 
must be dominated by magnetic processes.

The \textsl{Coronet}
 class~I objects with relatively well determined spectroscopic 
ages as young as 0.1~Myr are clearly detected as X-ray sources.
IRS\,7E is so far the only case of
a reliable X-ray detection of an object in earlier evolutionary
stage than class~I objects. 
The X-ray properties of this object are similar to those
of the class~I objects, showing that hot coronae and thus
magnetic activity exist already
in extremely young protostellar objects.

Concerning now the origin of the X-ray emission from young intermediate-mass 
stars, we first show that the X-ray emission from
HD~176386 originates not
from the Be star, but from a late type companion.
Then we demonstrate that the characteristics of the 
observed X-ray emission from the 
TY~CrA multiple system agree very well with the expected 
X-ray emission of the three late-type companions; there is no
need to assume that the Be star itself emits any X-rays.
The case of R~CrA remains unclear; its extremely hard
X-ray spectrum clearly suggests a magnetic origin of the emission.
One or several yet undiscovered and spatially unresolved
late-type companions may provide the
most straightforward explanation, although even in that case 
the extraordinarily high plasma temperature is very unusual.

Finally, we point out that
none of the numerous HH objects in the CrA star-forming region is detected
in the \textsl{Chandra} data despite the very high sensitivity. 
As X-ray emission at levels similar to those of the X-ray detected
HH objects in the other star-forming regions 
should have been easily detected in our data, this null-result
suggests that observable X-ray emission from HH objects
is not very frequent.

The further results can be summarized as follows: 
X-ray emission is detected from about half of the
brown dwarf candidates in the observed region.
Diffuse X-ray emission is tentatively detected in the central part
of the \textsl{Coronet} cluster, but its nature and origin remain unclear.


\begin{acknowledgements}
We are grateful to Lori Allen for providing us with the \textsl{Spitzer}
IRAC images of the CrA star-forming region and the classification of the sources
 prior to publication. 
We would like to thank Stefan Kraus for help with the \textsl{Spitzer}
images.
This work made extensive use of 
NASA's Astrophysics Data System Bibliographic Services
and the SIMBAD database (CDS, Strasbourg, France). This publication makes use of data products from the Two Micron All Sky Survey, which is a joint project of the University of Massachusetts and the Infrared Processing and Analysis Center/California Institute of Technology, funded by the National Aeronautics and Space Administration and the National Science Foundation.
\end{acknowledgements}

\bibliographystyle{aa}

\clearpage
\appendix
\section{Detailed source lists}

\begin{small}
\begin{longtable}{lrrrrrrrrrr}
\caption{\label{acis_extract.tab} Complete source list with parameters determined by \rm{\it acis\_extract} for the energy range of 0.5--8~keV, i.e. total counts, median photon energy, source significance, hardness ratios, as well as estimates of the observed flux. Also given are the ratio of the summed exposure map values at the source position relative to the maximum value, i.e., an information on the effective exposure time, and the number of observations in which a source was detected (N$_{\rm obs}$). This number can be lower than eight in the outer parts of the map. The manually added source is marked by an asterisk. }\\
\hline\hline
Src ID & Net        & Med. $E$ & Source      & HR1 & HR2 & HR3 & FLUX1  & FLUX2 & rel.exp.  \\
       & counts     & [keV]      & signif.     &     &     &	 & [cm$^{-2}$\, s$^{-1}$] &  [cm$^{-2}$\, s$^{-1}$] & (N$_{\rm obs}$)\\
\hline
\endfirsthead
\caption{cntd.}\\
\hline
\hline
Src ID & Net        & Med. $E$ & Source      & HR1 & HR2 & HR3 & FLUX1  & FLUX2 & rel.exp. \\
       & counts     & [keV]      & signif.     &     &     &     & [cm$^{-2}$\, s$^{-1}$] &  [cm$^{-2}$\, s$^{-1}$] & (N$_{\rm obs}$)\\
\hline
\endhead
\hline
\endfoot
190104.58-370129.6     & $46.2^{+9.6}_{-8.5}$	    & 1.3 &  4.8 & $	    -0.98_{-	  0.02}^{+	0.09}$ & $     -0.73_{-      0.10}^{+	   0.13}$ & $	  -3.05_{-	6.51}^{+      -NaN}$ &   1.99E-06 &   1.94E-06   & 0.50 (7)  \\
190105.05-370340.2     & $23.3^{+6.1}_{-5.0}$	    & 1.4 &  3.8 & $	    -0.64_{-	  0.17}^{+	0.21}$ & $     -0.42_{-      0.23}^{+	   0.24}$ & $	  -0.28_{-	0.39}^{+      0.38}$ &   1.41E-05 &   1.80E-05   & 0.02 (1)  \\
190108.31-365634.7     & $99.2^{+11.6}_{-10.6}$     & 2.8 &   8.5 & $	      0.27_{-	   0.10}^{+	 0.10}$ & $	 0.45_{-      0.14}^{+      0.14}$ & $      0.14_{-	 0.12}^{+      0.11}$ &   2.49E-06 &   3.35E-06  & 0.68 (7)  \\
190108.60-365721.3     & $3005.2^{+56.0}_{-55.0}$   & 1.8 &   53.7 & $        -0.17_{-      0.02}^{+	  0.02}$ & $	 -0.22_{-      0.02}^{+      0.02}$ & $     -0.03_{-	  0.02}^{+	0.02}$ &   7.24E-05 &	9.55E-05 & 0.72 (8)  \\
190111.30-365551.9     & $36.5^{+7.9}_{-6.8}$	    & 2.7 &  4.6 & $	     0.34_{-	  0.17}^{+	0.18}$ & $	0.45_{-      0.23}^{+	   0.24}$ & $	   0.07_{-	0.20}^{+      0.20}$ &   8.96E-07 &   1.32E-06   & 0.63 (7)  \\
190114.54-370102.3     & $21.1^{+6.5}_{-5.4}$	    & 2.9 &  3.3 & $	     0.37_{-	  0.23}^{+	0.24}$ & $	0.31_{-      0.39}^{+	   0.38}$ & $	   0.35_{-	0.25}^{+      0.26}$ &   4.62E-07 &   8.20E-07   & 0.59 (7)  \\
190115.32-365237.2     & $51.1^{+8.9}_{-7.8}$	    & 2.8 &  5.8 & $	     0.52_{-	  0.13}^{+	0.14}$ & $	0.63_{-      0.16}^{+	   0.20}$ & $	   0.09_{-	0.16}^{+      0.16}$ &   2.05E-06 &   2.54E-06   & 0.47 (6)  \\
190115.86-370344.3     & $862.5^{+30.6}_{-29.6}$    & 1.7 &  28.2 & $	     -0.21_{-	   0.03}^{+	 0.03}$ & $	-0.29_{-      0.04}^{+      0.04}$ & $     -0.04_{-	 0.05}^{+      0.05}$ &   2.87E-05 &   3.47E-05  & 0.58 (7)  \\
190116.26-365628.4     & $59.9^{+9.3}_{-8.2}$	    & 1.2 &  6.4 & $	    -0.85_{-	  0.07}^{+	0.10}$ & $     -0.59_{-      0.11}^{+	   0.12}$ & $	  -1.32_{-	0.15}^{+      -NaN}$ &   2.27E-06 &   1.76E-06   & 0.80 (8)  \\
190116.65-365638.9     & $28.3^{+7.1}_{-6.0}$	    & 3.5 &  4.0 & $	     0.75_{-	  0.13}^{+	0.18}$ & $	0.99_{-      0.30}^{+	   0.28}$ & $	   0.17_{-	0.21}^{+      0.21}$ &   6.67E-07 &   8.26E-07   & 0.81 (8)  \\
190117.52-365831.9     & $40.3^{+7.8}_{-6.8}$	    & 3.3 &  5.1 & $	      0.75_{-	   0.11}^{+	 0.14}$ & $	 1.08_{-      0.03}^{+      0.19}$ & $      0.26_{-	 0.17}^{+      0.17}$ &   9.85E-07 &   1.26E-06  & 0.75 (8)  \\
190118.29-365906.7     & $51.8^{+8.7}_{-7.6}$	    & 2.1 &  6.0 & $	      0.08_{-	   0.15}^{+	 0.15}$ & $	 0.26_{-      0.18}^{+      0.18}$ & $     -0.19_{-	 0.18}^{+      0.17}$ &   1.24E-06 &   1.56E-06  & 0.79 (8)  \\
190118.90-365828.4     & $156.8^{+13.8}_{-12.7}$    & 1.4 &  11.4 & $	      -0.53_{-      0.07}^{+	  0.07}$ & $	 -0.49_{-      0.08}^{+      0.08}$ & $     -0.30_{-	  0.14}^{+	0.14}$ &   3.62E-06 &	4.43E-06 & 0.81 (8)  \\
190119.39-370142.0     & $85.5^{+10.7}_{-9.6}$      & 1.8 &  8.0 & $	     -0.26_{-	   0.11}^{+	 0.11}$ & $	-0.12_{-      0.13}^{+      0.13}$ & $     -0.33_{-	 0.15}^{+      0.16}$ &   2.01E-06 &   2.76E-06  & 0.72 (8)  \\
190120.07-370422.9     & $28.8^{+7.4}_{-6.3}$	    & 3.0 &  3.9 & $	      0.44_{-	   0.18}^{+	 0.20}$ & $	 0.32_{-      0.30}^{+      0.30}$ & $      0.25_{-	 0.22}^{+      0.22}$ &   1.10E-06 &   1.21E-06  & 0.57 (7)  \\
190120.86-370302.9     & $166.4^{+14.2}_{-13.2}$    & 1.1 &  11.7 & $	      -0.86_{-      0.04}^{+	  0.05}$ & $	 -0.77_{-      0.05}^{+      0.06}$ & $     -0.73_{-	  0.15}^{+	0.21}$ &   7.35E-06 &	6.24E-06 & 0.63 (7)  \\
190122.40-370055.4     & $30.4^{+7.0}_{-5.9}$	    & 2.5 &  4.4 & $	      0.10_{-	   0.20}^{+	 0.20}$ & $	-0.51_{-      0.23}^{+      0.26}$ & $      0.52_{-	 0.23}^{+      0.26}$ &   6.57E-07 &   1.11E-06  & 0.66 (8)  \\
190123.95-365721.5     & $14.9^{+5.3}_{-4.2}$	    & 3.3 &  2.8 & $	      0.72_{-	   0.19}^{+	 0.27}$ & $	 0.32_{-      0.63}^{+      0.59}$ & $      0.64_{-	 0.23}^{+      0.29}$ &   2.70E-07 &   4.40E-07  & 0.81 (8)  \\
190125.61-370453.9     & $250.9^{+17.2}_{-16.2}$    & 1.1 &  14.6 & $	      -0.91_{-      0.03}^{+	  0.03}$ & $	 -0.70_{-      0.05}^{+      0.05}$ & $     -0.94_{-	  0.05}^{+	0.11}$ &   1.09E-05 &	9.58E-06 & 0.60 (7)  \\
190125.75-365919.3     & $2094.1^{+46.8}_{-45.8}$   & 1.6 &   44.7 & $     -0.38_{-  0.02}^{+	   0.02}$ & $	  -0.28_{-	0.02}^{+      0.02}$ & $     -0.31_{-	   0.03}^{+	 0.03}$ &   4.40E-05 &   6.13E-05        & 0.81 (8)  \\
190126.10-365501.9     & $90.2^{+10.8}_{-9.7}$      & 2.4 &  8.4 & $	      0.21_{-	   0.11}^{+	 0.11}$ & $	 0.20_{-      0.15}^{+      0.15}$ & $      0.10_{-	 0.13}^{+      0.13}$ &   2.57E-06 &   2.48E-06  & 0.85 (8)  \\
190127.15-365908.6     & $883.3^{+30.8}_{-29.8}$    & 1.0 &  28.7 & $	      -0.85_{-      0.02}^{+	  0.02}$ & $	 -0.80_{-      0.02}^{+      0.02}$ & $     -0.49_{-	  0.09}^{+	0.09}$ &   3.33E-05 &	2.61E-05 & 0.82 (8)  \\
190128.72-365931.9     & $3700.5^{+61.9}_{-60.8}$   & 1.5 &   59.8 & $     -0.50_{-  0.01}^{+	   0.01}$ & $	  -0.42_{-	0.02}^{+      0.02}$ & $     -0.32_{-	   0.03}^{+	 0.03}$ &   7.52E-05 &   1.02E-04        & 0.87 (8)  \\
190129.01-370148.8     & $969.4^{+32.2}_{-31.2}$    & 1.1 &  30.1 & $	      -0.81_{-      0.02}^{+	  0.02}$ & $	 -0.75_{-      0.02}^{+      0.02}$ & $     -0.45_{-	  0.07}^{+	0.08}$ &   3.19E-05 &	2.69E-05 & 0.85 (8)  \\
190131.28-365931.0     & $13.0^{+5.0}_{-3.8}$	    & 3.0 &  2.6 & $	      0.66_{-	   0.22}^{+	 0.30}$ & $	 0.87_{-      0.16}^{+      0.55}$ & $      0.27_{-	 0.32}^{+      0.32}$ &   2.79E-07 &   3.43E-07  & 0.94 (8)  \\
190131.73-365445.4     & $6.2^{+4.1}_{-2.9}$	    & 1.0 & 1.5 & $ -0.98_{-	  0.18}^{+	0.46}$ & $     -0.82_{-      0.20}^{+	   0.49}$ & $	  -0.14_{-	1.35}^{+      0.99}$ &   3.36E-07 &   2.01E-07           & 0.75 (8)  \\
190132.01-365121.7     & $26.4^{+7.3}_{-6.2}$	    & 3.2 &  3.6 & $	      0.62_{-	   0.17}^{+	 0.20}$ & $	 0.34_{-      0.33}^{+      0.33}$ & $      0.31_{-	 0.23}^{+      0.23}$ &   7.21E-07 &   7.84E-07  & 0.72 (8)  \\
190132.34-365803.1     & $22.8^{+6.0}_{-4.9}$	    & 1.6 &  3.8 & $	     -0.81_{-	   0.12}^{+	 0.20}$ & $	-0.37_{-      0.23}^{+      0.23}$ & $     -0.68_{-	 0.28}^{+      0.40}$ &   3.69E-07 &   6.28E-07  & 0.89 (8)  \\
190133.61-370606.2     & $29.0^{+7.5}_{-6.4}$	    & 2.0 &  3.9 & $	     -0.08_{-	   0.21}^{+	 0.20}$ & $	-0.23_{-      0.24}^{+      0.24}$ & $     -0.05_{-	 0.30}^{+      0.28}$ &   4.78E-07 &   1.16E-06  & 0.57 (7)  \\
190133.84-365745.0     & $841.9^{+30.0}_{-29.0}$    & 2.1 &  28.0 & $	       0.08_{-      0.04}^{+	  0.04}$ & $	  0.13_{-      0.04}^{+      0.04}$ & $     -0.07_{-	  0.04}^{+	0.04}$ &   1.61E-05 &	2.21E-05 & 0.94 (8)  \\
190134.65-365438.4     & $8.3^{+4.4}_{-3.3}$	    & 2.6 &  1.9 & $	      0.24_{-	   0.41}^{+	 0.40}$ & $	-0.37_{-      0.53}^{+      0.53}$ & $      0.39_{-	 0.51}^{+      0.52}$ &   3.08E-07 &   2.25E-07  & 0.89 (8)  \\
190134.84-370056.7     & $25457.3^{+160.6}_{-159.6}$& 1.3 &   158.5 & $     -0.56_{-	      0.01}^{+      0.01}$ & $     -0.56_{-	 0.01}^{+      0.01}$ & $     -0.24_{-      0.01}^{+	  0.01}$ &   6.03E-04 &6.89E-04  & 0.89 (8)  \\
190135.42-370340.9     & $41.3^{+7.9}_{-6.8}$	    & 1.8 &  5.2 & $	     -0.26_{-	   0.17}^{+	 0.17}$ & $	-0.24_{-      0.20}^{+      0.20}$ & $      0.03_{-	 0.24}^{+      0.23}$ &   7.58E-07 &   1.18E-06  & 0.84 (8)  \\
190137.07-365601.9     & $114.0^{+11.8}_{-10.7}$    & 3.8 &   9.7 & $	       0.83_{-      0.05}^{+	  0.07}$ & $	  0.58_{-      0.19}^{+      0.22}$ & $      0.67_{-	  0.07}^{+	0.08}$ &   3.54E-06 &	3.37E-06 & 0.79 (8)  \\
190139.15-365329.6     & $7715.4^{+88.9}_{-87.9}$   & 1.1 &   86.8 & $     -0.79_{-  0.01}^{+	   0.01}$ & $	  -0.73_{-	0.01}^{+      0.01}$ & $     -0.44_{-	   0.02}^{+	 0.02}$ &   2.19E-04 &   2.17E-04        & 0.85 (8)  \\
190139.34-370207.8     & $710.8^{+27.7}_{-26.7}$    & 1.0 &  25.6 & $	      -0.89_{-      0.02}^{+	  0.02}$ & $	 -0.84_{-      0.02}^{+      0.02}$ & $     -0.63_{-	  0.10}^{+	0.11}$ &   2.37E-05 &	1.96E-05 & 0.90 (8)  \\
190140.40-365142.4     & $586.8^{+25.5}_{-24.4}$    & 2.6 &  23.1 & $	       0.41_{-      0.04}^{+	  0.04}$ & $	  0.39_{-      0.05}^{+      0.05}$ & $      0.03_{-	  0.05}^{+	0.05}$ &   1.33E-05 &	1.70E-05 & 0.80 (8)  \\
190140.63-365758.9     & $2.4^{+2.9}_{-1.6}$	    & 0.7 &    0.8 & $     -1.35_{-  0.65}^{+		0.94}$ & $     -1.08_{-      0.11}^{+	   0.82}$ & $	   0.52_{-	0.00}^{+      2.14}$ &   3.35E-07 &   1.03E-07   & 0.68 (8)  \\
190140.67-365923.9     & $13.3^{+4.8}_{-3.7}$	    & 3.5 &  2.7 & $	      0.58_{-	   0.25}^{+	 0.30}$ & $	-1.09_{-      0.12}^{+      0.84}$ & $      1.02_{-	 0.01}^{+      0.30}$ &   2.42E-07 &   3.61E-07  & 0.90 (8)  \\
190140.70-370052.9     & $70.8^{+9.5}_{-8.5}$	    & 2.8 &  7.4 & $	      0.19_{-	   0.13}^{+	 0.13}$ & $	-0.11_{-      0.19}^{+      0.18}$ & $      0.38_{-	 0.14}^{+      0.15}$ &   1.39E-06 &   1.84E-06  & 0.95 (8)  \\
190140.81-365234.0     & $21015.9^{+146.0}_{-145.0}$& 1.4 &   143.9 & $     -0.57_{-	      0.01}^{+      0.01}$ & $     -0.52_{-	 0.01}^{+      0.01}$ & $     -0.31_{-      0.01}^{+	  0.01}$ &   4.87E-04 & 6.18E-04 & 0.80 (8)  \\
190141.35-370419.3     & $29.2^{+7.1}_{-6.0}$	    & 1.9 &  4.1 & $	     -0.21_{-	   0.20}^{+	 0.20}$ & $	-0.52_{-      0.21}^{+      0.24}$ & $      0.32_{-	 0.30}^{+      0.30}$ &   5.29E-07 &   8.10E-07  & 0.85 (8)  \\
190141.49-370441.7     & $32.0^{+7.4}_{-6.3}$	    & 2.1 &  4.3 & $	      0.05_{-	   0.20}^{+	 0.19}$ & $	-0.31_{-      0.23}^{+      0.23}$ & $      0.11_{-	 0.28}^{+      0.27}$ &   4.64E-07 &   9.07E-07  & 0.81 (8)  \\
190141.55-365831.6     & $3037.4^{+56.1}_{-55.1}$   & 2.7 &   54.1 & $      0.40_{-  0.02}^{+	   0.02}$ & $	   0.38_{-	0.02}^{+      0.02}$ & $      0.16_{-	   0.02}^{+	 0.02}$ &   9.61E-05 &   1.14E-04        & 0.65 (8)  \\
190141.62-365953.1     & $1038.2^{+33.2}_{-32.2}$   & 1.3 &   31.2 & $     -0.61_{-  0.02}^{+	   0.03}$ & $	  -0.60_{-	0.03}^{+      0.03}$ & $     -0.25_{-	   0.06}^{+	 0.06}$ &   2.28E-05 &   2.87E-05        & 0.89 (8)  \\
190143.12-365020.9     & $45.2^{+8.8}_{-7.7}$	    & 3.2 &  5.1 & $	      0.71_{-	   0.11}^{+	 0.14}$ & $	 0.81_{-      0.15}^{+      0.26}$ & $      0.39_{-	 0.15}^{+      0.16}$ &   9.41E-07 &   1.44E-06  & 0.71 (8)  \\
190143.80-370614.0     & $155.5^{+14.1}_{-13.0}$    & 2.2 &  11.0 & $	       0.07_{-      0.08}^{+	  0.08}$ & $	 -0.04_{-      0.11}^{+      0.10}$ & $      0.02_{-	  0.11}^{+	0.11}$ &   3.81E-06 &	4.75E-06 & 0.77 (8)  \\
190144.22-365853.6     & $72.5^{+9.6}_{-8.5}$	    & 2.8 &  7.6 & $	      0.34_{-	   0.12}^{+	 0.12}$ & $	 0.46_{-      0.16}^{+      0.17}$ & $      0.14_{-	 0.14}^{+      0.14}$ &   2.08E-06 &   2.56E-06  & 0.69 (8)  \\
190145.91-364929.3     & $119.6^{+12.8}_{-11.7}$    & 3.5 &   9.4 & $	       0.76_{-      0.06}^{+	  0.07}$ & $	  0.64_{-      0.15}^{+      0.18}$ & $      0.57_{-	  0.08}^{+	0.09}$ &   3.82E-06 &	4.45E-06 & 0.58 (7)  \\
190146.18-370241.8     & $20.7^{+5.9}_{-4.8}$	    & 3.0 &  3.5 & $	      0.59_{-	   0.20}^{+	 0.23}$ & $	 0.71_{-      0.23}^{+      0.35}$ & $      0.11_{-	 0.26}^{+      0.25}$ &   5.74E-07 &   5.83E-07  & 0.83 (8)  \\
190147.01-365326.9     & $22.9^{+6.2}_{-5.1}$	    & 3.7 &  3.7 & $	  0.74_{-      0.15}^{+      0.20}$ & $      0.04_{-	  0.61}^{+	0.53}$ & $	0.77_{-      0.15}^{+	   0.21}$ &   4.84E-07 &   6.83E-07      & 0.79 (8)  \\
190148.02-365722.4     & $3604.2^{+61.0}_{-60.0}$   & 3.3 &   59.0 & $      0.72_{-	 0.01}^{+      0.01}$ & $      0.63_{-      0.02}^{+	  0.02}$ & $	  0.40_{-      0.02}^{+      0.02}$ &	8.42E-05 &   9.26E-05    & 0.97 (8)  \\
190148.46-365714.5     & $21.2^{+5.8}_{-4.7}$	    & 4.1 &  3.7 & $	  1.02_{-      0.01}^{+      0.15}$ & $      1.18_{-	  0.48}^{+	1.12}$ & $	0.83_{-      0.12}^{+	   0.20}$ &   4.77E-07 &   5.43E-07      & 0.97 (8)  \\
190148.53-370611.2     & $50.6^{+9.1}_{-8.0}$	    & 4.1 &  5.6 & $	  0.70_{-      0.10}^{+      0.13}$ & $      0.54_{-	  0.26}^{+	0.30}$ & $	0.58_{-      0.13}^{+	   0.14}$ &   2.44E-06 &   1.56E-06      & 0.72 (8)  \\
190148.97-370151.7     & $8.0^{+4.1}_{-2.9}$	    & 2.8 & 1.9 & $	 0.82_{-      0.18}^{+      0.40}$ & $      0.56_{-	 0.51}^{+      0.63}$ & $      0.22_{-      0.45}^{+	  0.42}$ &   2.75E-07 &   2.85E-07       & 0.68 (8)  \\
190149.35-370028.6     & $110.3^{+11.6}_{-10.5}$    & 0.9 &   9.5 & $	  -0.94_{-	0.03}^{+      0.05}$ & $     -0.87_{-	   0.05}^{+	 0.06}$ & $	-0.86_{-      0.16}^{+      0.42}$ &   5.04E-06 &   3.11E-06     & 0.89 (8)  \\
190150.38-365935.3     & $10.4^{+4.4}_{-3.3}$	    & 4.1 &  2.4 & $	  0.84_{-      0.15}^{+      0.33}$ & $      1.10_{-	  0.13}^{+	0.82}$ & $	0.45_{-      0.32}^{+	   0.35}$ &   2.80E-07 &   2.96E-07      & 0.89 (8)  \\
190150.45-365638.1     & $259.2^{+17.1}_{-16.1}$    & 2.1 &  15.1 & $	   0.04_{-	0.06}^{+      0.06}$ & $     -0.03_{-	   0.08}^{+	 0.08}$ & $	 0.05_{-      0.08}^{+      0.08}$ &   5.18E-06 &   6.83E-06     & 0.95 (8)  \\
190150.66-365809.9     & $2622.5^{+52.2}_{-51.2}$   & 3.1 &   50.2 & $      0.57_{-	 0.02}^{+      0.02}$ & $      0.49_{-      0.03}^{+	  0.03}$ & $	  0.29_{-      0.02}^{+      0.02}$ &	8.70E-05 &   9.74E-05    & 0.68 (8)  \\
190151.11-365412.5     & $59.0^{+8.9}_{-7.8}$	    & 1.9 &  6.7 & $	 -0.10_{-      0.14}^{+      0.14}$ & $      0.05_{-	  0.16}^{+	0.16}$ & $     -0.27_{-      0.18}^{+	   0.18}$ &   1.04E-06 &   1.63E-06      & 0.90 (8)  \\
190152.10-370542.4     & $68.3^{+10.0}_{-8.9}$      & 3.4 &  6.8 & $	  0.69_{-      0.09}^{+      0.11}$ & $      0.40_{-	  0.23}^{+	0.24}$ & $	0.55_{-      0.11}^{+	   0.12}$ &   1.20E-06 &   1.99E-06      & 0.79 (8)  \\
190152.20-365809.0     & $15.4^{+5.1}_{-4.0}$	    & 5.1 &  3.0 & $	  1.02_{-      0.01}^{+      0.21}$ & $     -0.08_{-	  0.00}^{+	1.60}$ & $	1.01_{-      0.00}^{+	   0.21}$ &   5.60E-07 &   4.86E-07      & 0.81 (8)  \\
190152.63-365700.2     & $247.1^{+16.8}_{-15.7}$    & 2.2 &  14.7 & $	   0.13_{-	0.07}^{+      0.07}$ & $      0.29_{-	   0.08}^{+	 0.08}$ & $	-0.16_{-      0.08}^{+      0.08}$ &   4.99E-06 &   6.45E-06     & 0.97 (8)  \\
190153.67-365708.3     & $980.1^{+32.3}_{-31.3}$    & 3.3 &  30.3 & $	   0.50_{-	0.03}^{+      0.03}$ & $      0.18_{-	   0.05}^{+	 0.05}$ & $	 0.43_{-      0.03}^{+      0.03}$ &   2.34E-05 &   2.59E-05     & 0.95 (8)  \\
190153.92-365228.8     & $23.1^{+6.5}_{-5.4}$	    & 3.5 &  3.6 & $	  0.73_{-      0.15}^{+      0.20}$ & $      0.84_{-	  0.18}^{+	0.46}$ & $	0.48_{-      0.21}^{+	   0.22}$ &   6.05E-07 &   7.05E-07      & 0.76 (8)  \\
190155.31-365722.0     & $253.1^{+17.0}_{-15.9}$    & 5.2 &  14.9 & $	   0.97_{-	0.01}^{+      0.02}$ & $      0.78_{-	   0.22}^{+	 0.41}$ & $	 0.95_{-      0.02}^{+      0.03}$ &   1.00E-05 &   7.54E-06     & 0.85 (8)  \\
190155.61-365651.1     & $40.1^{+7.5}_{-6.4}$	    & 4.5 &  5.4 & $	  0.96_{-      0.03}^{+      0.10}$ & $      0.65_{-	  0.38}^{+	0.56}$ & $	0.81_{-      0.10}^{+	   0.14}$ &   1.56E-06 &   1.24E-06      & 0.80 (8)  \\
190155.76-365727.7     & $19.3^{+5.6}_{-4.4}$	    & 4.4 &  3.5 & $	  1.02_{-      0.00}^{+      0.17}$ & $      1.06_{-	  0.06}^{+	0.67}$ & $	0.60_{-      0.20}^{+	   0.24}$ &   7.67E-07 &   7.15E-07      & 0.68 (8)  \\
190155.85-365204.3     & $371.7^{+20.5}_{-19.5}$    & 3.7 &  18.1 & $	   0.78_{-	0.03}^{+      0.04}$ & $      0.53_{-	   0.09}^{+	 0.10}$ & $	 0.61_{-      0.04}^{+      0.05}$ &   1.14E-05 &   1.12E-05     & 0.80 (8)  \\
190156.39-365728.4     & $106.4^{+11.4}_{-10.3}$    & 4.7 &   9.4 & $	   0.98_{-	0.01}^{+      0.04}$ & $      0.36_{-	   0.73}^{+	 0.69}$ & $	 0.96_{-      0.02}^{+      0.05}$ &   6.32E-06 &   5.01E-06     & 0.55 (8)  \\
190157.46-370311.9     & $24.2^{+6.4}_{-5.3}$	    & 1.4 &  3.8 & $	 -0.47_{-      0.20}^{+      0.22}$ & $     -0.48_{-	  0.22}^{+	0.24}$ & $     -0.15_{-      0.40}^{+	   0.38}$ &   4.04E-07 &   6.64E-07      & 0.85 (8)  \\
190158.32-370027.5     & $23.1^{+6.0}_{-4.9}$	    & 1.8 &  3.9 & $	 -0.19_{-      0.23}^{+      0.23}$ & $     -0.22_{-	  0.27}^{+	0.26}$ & $     -0.11_{-      0.34}^{+	   0.32}$ &   4.48E-07 &   6.49E-07      & 0.89 (8)  \\
190158.79-365750.1     & $13.3^{+4.8}_{-3.7}$	    & 1.3 &  2.7 & $	 -0.78_{-      0.17}^{+      0.29}$ & $     -0.58_{-	  0.24}^{+	0.29}$ & $     -1.37_{-      0.71}^{+	   0.95}$ &   3.11E-07 &   5.39E-07      & 0.63 (7)  \\
190200.11-370222.3     & $524.4^{+24.0}_{-22.9}$    & 1.0 &  21.9 & $	  -0.92_{-	0.02}^{+      0.02}$ & $     -0.86_{-	   0.02}^{+	 0.03}$ & $	-0.55_{-      0.13}^{+      0.15}$ &   2.05E-05 &   1.47E-05     & 0.87 (8)  \\
190200.49-365507.2     & $25.1^{+6.3}_{-5.2}$	    & 3.4 &  4.0 & $	  0.96_{-      0.04}^{+      0.15}$ & $      0.79_{-	  0.25}^{+	0.53}$ & $	0.62_{-      0.17}^{+	   0.21}$ &   7.74E-07 &   6.92E-07      & 0.88 (8)  \\
190201.92-370743.0     & $5893.5^{+77.8}_{-76.8}$   & 1.1 &   75.7 & $     -0.78_{-	 0.01}^{+      0.01}$ & $     -0.74_{-      0.01}^{+	  0.01}$ & $	 -0.39_{-      0.03}^{+      0.03}$ &	5.79E-04 &   4.93E-04    & 0.20 (2)  \\
190201.94-365400.1     & $42.6^{+7.8}_{-6.8}$	    & 3.4 &  5.4 & $	      0.76_{-	   0.10}^{+	 0.13}$ & $	 0.56_{-      0.24}^{+      0.28}$ & $      0.42_{-	 0.16}^{+      0.17}$ &   1.08E-06 &   1.18E-06  & 0.88 (8)  \\
190202.16-365919.1     & $11.9^{+4.7}_{-3.6}$	    & 4.6 &  2.5 & $	      0.88_{-	   0.11}^{+	 0.30}$ & $	 1.97_{-      4.13}^{+      2.37}$ & $      0.87_{-	 0.13}^{+      0.29}$ &   3.65E-07 &   3.18E-07  & 0.93 (8)  \\
190203.82-365851.0     & $39.8^{+7.5}_{-6.4}$	    & 4.1 &  5.3 & $	      0.92_{-	   0.06}^{+	 0.12}$ & $	 0.55_{-      0.35}^{+      0.42}$ & $      0.70_{-	 0.12}^{+      0.15}$ &   1.22E-06 &   1.08E-06  & 0.92 (8)  \\
190204.22-370420.9     & $175.6^{+14.6}_{-13.5}$    & 2.3 &  12.1 & $	       0.16_{-      0.08}^{+	  0.08}$ & $	  0.01_{-      0.10}^{+      0.10}$ & $      0.13_{-	  0.09}^{+	0.09}$ &   4.10E-06 &	5.40E-06 & 0.75 (8)  \\
190210.72-370559.2     & $42.7^{+8.3}_{-7.2}$	    & 2.6 &  5.2 & $	      0.10_{-	   0.17}^{+	 0.17}$ & $	-0.47_{-      0.20}^{+      0.22}$ & $      0.53_{-	 0.18}^{+      0.21}$ &   2.67E-06 &   2.50E-06  & 0.38 (3)  \\
190211.99-370309.4     & $77.1^{+10.3}_{-9.2}$      & 1.2 &  7.5 & $	     -0.76_{-	   0.08}^{+	 0.09}$ & $	-0.79_{-      0.07}^{+      0.10}$ & $     -0.07_{-	 0.32}^{+      0.30}$ &   2.87E-06 &   2.30E-06  & 0.76 (8)  \\
190215.00-365232.2     & $30.4^{+7.3}_{-6.2}$	    & 1.8 &  4.2 & $	     -0.29_{-	   0.19}^{+	 0.20}$ & $	-0.75_{-      0.16}^{+      0.23}$ & $      0.63_{-	 0.23}^{+      0.30}$ &   1.19E-06 &   1.48E-06  & 0.48 (7)  \\
190216.54-365637.8     & $27.1^{+6.9}_{-5.8}$	    & 3.7 &  3.9 & $	      0.84_{-	   0.10}^{+	 0.17}$ & $	 0.43_{-      0.48}^{+      0.51}$ & $      0.71_{-	 0.15}^{+      0.19}$ &   6.36E-07 &   7.70E-07  & 0.84 (8)  \\
190219.19-370346.3     & $53.5^{+9.2}_{-8.2}$	    & 3.4 &  5.8 & $	      0.77_{-	   0.09}^{+	 0.12}$ & $	 0.72_{-      0.19}^{+      0.28}$ & $      0.51_{-	 0.13}^{+      0.14}$ &   1.31E-06 &   1.91E-06  & 0.66 (8)  \\
190221.22-365013.5     & $52.0^{+9.0}_{-7.9}$	    & 2.7 &  5.8 & $	      0.24_{-	   0.15}^{+	 0.15}$ & $	-0.11_{-      0.21}^{+      0.20}$ & $      0.26_{-	 0.18}^{+      0.18}$ &   4.18E-06 &   4.32E-06  & 0.26 (2)  \\
190221.79-365604.2     & $59.5^{+9.6}_{-8.5}$	    & 4.4 &  6.2 & $	      0.85_{-	   0.07}^{+	 0.10}$ & $	 0.03_{-      0.37}^{+      0.34}$ & $      0.79_{-	 0.09}^{+      0.11}$ &   1.94E-06 &   1.78E-06  & 0.80 (8)  \\
190222.12-365313.2     & $20.0^{+6.4}_{-5.3}$	    & 1.9 &  3.1 & $	     -0.24_{-	   0.25}^{+	 0.25}$ & $	-0.02_{-      0.28}^{+      0.27}$ & $     -0.55_{-	 0.28}^{+      0.33}$ &   7.58E-07 &   1.22E-06  & 0.39 (3)  \\
190222.13-365541.0     & $5458.3^{+75.0}_{-73.9}$   & 1.3 &   72.8 & $     -0.63_{-  0.01}^{+	   0.01}$ & $	  -0.55_{-	0.01}^{+      0.01}$ & $     -0.38_{-	   0.02}^{+	 0.02}$ &   1.89E-04 &   2.25E-04        & 0.59 (8)  \\
190222.71-365308.5     & $21.5^{+6.4}_{-5.3}$	    & 2.5 &  3.4 & $	      0.17_{-	   0.24}^{+	 0.24}$ & $	 0.11_{-      0.34}^{+      0.32}$ & $      0.16_{-	 0.29}^{+      0.28}$ &   1.26E-06 &   1.46E-06  & 0.35 (3)  \\
190227.05-365813.2     & $359.7^{+20.4}_{-19.4}$    & 1.6 &  17.6 & $	      -0.43_{-      0.05}^{+	  0.05}$ & $	 -0.37_{-      0.06}^{+      0.06}$ & $     -0.20_{-	  0.08}^{+	0.08}$ &   8.68E-06 &	1.15E-05 & 0.71 (8)  \\
190233.07-365821.1$^*$ & $104.6^{+11.5}_{-10.4}$    & 2.2 &   9.1 & $	       0.01_{-      0.10}^{+	  0.10}$ & $	 -0.48_{-      0.12}^{+      0.13}$ & $      0.50_{-	  0.12}^{+	0.13}$ &   8.05E-06 &	8.62E-06 & 0.28 (2)  \\
\hline  																											        
\end{longtable}
\end{small}

\begin{table*}
\begin{centering}
\caption{X-ray spectral fitting results for one- and two-temperature 
models. The fitting parameters are the absorbing hydrogen column density 
$N_{\rm H}$, the plasma temperatures $T$, and the normalization factors from XSPEC
(defined as $10^{-14}/(4 \pi D^2) \int n_{\rm e} n_{\rm H} dV$). We also list
the extinction-corrected (unabsorbed) X-ray luminosity $L_{\rm X}$
(integrated over the 0.2--8 keV band) derived from the model parameters,
the goodness-of-fit measure $\chi^2_{\rm red}$, and the number of degrees of freedom (DOF).
The fitting parameters for R~CrA are given in the text.
}
\label{table_fit}       
\begin{tabular}{lrrrrrrrr}
\hline\noalign{\smallskip}
Source                    & $N_{\rm H}$                    & $T_1$                       & $T_2$                       & norm$_1$                           &  norm$_2$                         &  $L_{\rm X}$            & $\chi^2_{\rm red}$  & DOF \\
                          & [10$^{22}$ cm$^{-2}$]          & [MK]                    & [MK]                    &                                       &                                       & [erg s$^{-1}$]  & & \\
\hline\noalign{\smallskip}

190108.60--365721.3 	  & 0.72       $^{+0.02}_{-0.02}$  &  6.8 $^{+0.5}_{-0.5}$    & 49.4 $^{+3.1}_{-2.3}$    & $(3.10^{+0.08}_{-0.08}) \, 10^{-4}$ & $(1.67^{+0.16}_{-0.16}) \, 10^{-4}$ & $1.23 \times 10^{30}$ & 0.93 & 64 \\\vspace{1mm}

190115.86--370344.3 	  & 0.22       $^{+0.03}_{-0.03}$  & 77.5 $^{+17.3}_{-11.5}$  & 			 & $(8.80^{+0.35}_{-0.35}) \, 10^{-5}$ &					 & $2.82 \times 10^{29}$ & 0.50 & 19 \\\vspace{1mm}


190118.90--365828.4 	  & 0.14       $^{+0.06}_{-0.05}$  & 27.6 $^{+6.3}_{-5.3}$    & 			 & $(1.18^{+0.12}_{-0.12}) \, 10^{-5}$ &					 & $2.70 \times 10^{28}$ & 1.09 & 5\\\vspace{1mm}

190119.39--370142.0 	  & 0.54       $^{+0.14}_{-0.11}$  & 33.1 $^{+10.1}_{-7.0}$   & 			 & $(1.17^{+0.15}_{-0.15}) \, 10^{-5}$ &					  & $2.87 \times 10^{28}$ & 0.08 & 1\\\vspace{1mm}

190120.86--370302.9 	  & 0  $^{+0.03}_{     }$  & 14.9 $^{+2.2}_{-1.0}$    & 			 & $(1.24^{+0.12}_{-0.12}) \, 10^{-5}$ &					  & $2.49 \times 10^{28}$ & 1.80 & 5\\\vspace{1mm}

190125.61--370453.9 	  & 0.32       $^{+0.03}_{-0.03}$  &  7.9 $^{+0.5}_{-0.3}$    & 			 & $(4.06^{+0.31}_{-0.31}) \, 10^{-5}$ &					  & $8.79 \times 10^{28}$ & 1.96 & 5\\\vspace{1mm}

190125.75--365919.3 	  & 1.07       $^{+0.02}_{-0.02}$  &  7.2 $^{+0.2}_{-0.2}$    & 25.6 $^{+1.2}_{-1.2}$    & $(4.03^{+0.23}_{-0.23}) \, 10^{-4}$ & $(2.39^{+0.09}_{-0.09}) \, 10^{-4}$  & $1.37 \times 10^{30}$ & 0.57 & 45\\\vspace{1mm}

190127.15--365908.6 	  & 0.04       $^{+0.01}_{-0.01}$  &  3.4 $^{+0.2}_{-0.2}$    & 13.8 $^{+0.6}_{-0.6}$    & $(3.52^{+0.45}_{-0.45}) \, 10^{-5}$ & $(4.26^{+0.21}_{-0.21}) \, 10^{-5}$  & $1.33 \times 10^{29}$ & 1.22 & 17\\\vspace{1mm}

190128.72--365931.9 	  & 0.53       $^{+0.01}_{-0.01}$  &  8.8 $^{+0.5}_{-0.5}$    & 24.6 $^{+0.8}_{-0.7}$    & $(3.30^{+0.08}_{-0.08}) \, 10^{-4}$ & $(1.41^{+0.09}_{-0.09}) \, 10^{-4}$  & $1.03 \times 10^{30}$ & 0.75 & 78\\\vspace{1mm}

190129.01--370148.8 	  & 0.15       $^{+0.01}_{-0.01}$  &  3.5 $^{+0.2}_{-0.2}$    & 14.2 $^{+0.6}_{-0.6}$    & $(6.60^{+0.65}_{-0.65}) \, 10^{-5}$ & $(4.95^{+0.26}_{-0.26}) \, 10^{-5}$  & $1.89 \times 10^{29}$ & 0.93 & 19\\\vspace{1mm}

190133.84--365745.0 	  & 1.63       $^{+0.06}_{-0.06}$  & 26.1 $^{+1.3}_{-1.3}$    & 			 & $(1.91^{+0.08}_{-0.08}) \, 10^{-4}$ &					   & $4.28 \times 10^{29}$ & 0.57 & 19\\\vspace{1mm}

190134.84--370056.7	  & 0.137      $^{+0.003}_{-0.003}$& 11.3 $^{+0.1}_{-0.1}$    & 40.6 $^{+1.0}_{-1.0}$    & $(6.23^{+0.10}_{-0.10}) \, 10^{-4}$ & $(1.07^{+0.01}_{-0.01}) \, 10^{-3}$ & $4.17 \times 10^{30}$ & 1.39 & 213\\\vspace{1mm}

190139.15--365329.6	  & 0.07       $^{+0.01}_{-0.01}$  &  9.3 $^{+0.1}_{-0.1}$    & 21.5 $^{+0.8}_{-0.8}$    & $(2.41^{+0.06}_{-0.06}) \, 10^{-4}$ & $(2.68^{+0.07}_{-0.07}) \, 10^{-4}$ & $1.08 \times 10^{30}$ & 1.18 & 99\\\vspace{1mm}

190139.34--370207.8	  & 0.457      $^{+0.02}_{-0.02}$  &  2.82 $^{+0.1}_{-0.1}$   & 13.6 $^{+1.3}_{-1.3}$    & $(3.03^{+0.30}_{-0.30}) \, 10^{-5}$ & $(4.50^{+0.29}_{-0.29}) \, 10^{-4}$ & $6.24 \times 10^{29}$ & 0.65 & 13\\\vspace{1mm}

190140.40--365142.4	  & 2.34       $^{+0.11}_{-0.10}$  & 29.5 $^{+1.5}_{-1.5}$    & 			 & $(1.78^{+0.09}_{-0.09}) \, 10^{-4}$ &					   & $4.19 \times 10^{29}$ & 0.65 & 12\\\vspace{1mm}

190140.81--365234.0	  & 0.357      $^{+0.004}_{-0.004}$& 10.2 $^{+0.1}_{-0.1}$    & 27.6 $^{+0.5}_{-0.5}$    & $(1.37^{+0.02}_{-0.02}) \, 10^{-3}$ & $(8.68^{+0.16}_{-0.16}) \, 10^{-4}$ & $5.01 \times 10^{30}$ & 1.13 & 186\\\vspace{1mm}

190141.55--365831.6$^{\bf [1]}$    & 2.42       $^{+0.05}_{-0.05}$  & 33.8 $^{+0.9}_{-0.9}$    & 			 & $(1.02^{+0.02}_{-0.02}) \, 10^{-3}$ &					   & $2.88 \times 10^{30}$ & 1.07 & 72\\\vspace{1mm}

190141.62--365953.1	  & 0.32       $^{+0.02}_{-0.02}$  &  9.5 $^{+0.7}_{-0.6}$    & 28.5 $^{+2.4}_{-2.4}$    & $(3.98^{+0.33}_{-0.33}) \, 10^{-5}$ & $(5.81^{+0.33}_{-0.33}) \, 10^{-4}$  & $2.21 \times 10^{29}$ & 0.72 & 21\\\vspace{1mm}

190148.02--365722.4$^{\bf [2]}$    & 3.87       $^{+0.07}_{-0.07}$  & 42.8 $^{+1.2}_{-1.2}$    & 			 & $(1.13^{+0.02}_{-0.02}) \, 10^{-3}$ &					   & $3.09 \times 10^{30}$ & 0.90 & 84\\\vspace{1mm}

190149.35--370028.6	  & 0.24       $^{+0.04}_{-0.04}$  &  4.6 $^{+0.3}_{-0.3}$    & 			 & $(1.91^{+0.24}_{-0.23}) \, 10^{-5}$ &					   & $3.07 \times 10^{28}$ & 1.39 & 3\\\vspace{1mm}

190150.45--365638.1	  & 1.09       $^{+0.10}_{-0.09}$  & 37.4 $^{+4.8}_{-4.3}$    & 			 & $(3.86^{+0.28}_{-0.28}) \, 10^{-5}$ &					    & $9.97 \times 10^{28}$ & 0.69 & 5\\\vspace{1mm}

190150.66--365809.9$^{\bf [1]}$    & 2.75       $^{+0.07}_{-0.06}$  & 48.2 $^{+1.6}_{-1.6}$    & 			 & $(7.90^{+0.18}_{-0.18}) \, 10^{-4}$ &					   & $2.45 \times 10^{30}$ & 1.41 & 64\\\vspace{1mm}


190152.63--365700.2	  & 1.81       $^{+0.13}_{-0.11}$  & 24.3 $^{+2.1}_{-1.9}$    & 			 & $(6.39^{+0.48}_{-0.48}) \, 10^{-5}$ &					   & $1.39 \times 10^{29}$ & 1.11 & 5\\\vspace{1mm}



190155.31--365722.0        &27.67       $^{+1.66}_{-1.47}$  & 99.3 $^{+25.1}_{-16.4}$  & 			 & $(3.30^{+0.24}_{-0.24}) \, 10^{-4}$ &					    & $1.10 \times 10^{30}$ & 0.82 & 5\\\vspace{1mm}

190155.85--365204.3        & 3.40       $^{+0.30}_{-0.26}$  & $>740^{    }_{-169}$     &                          & $(9.84^{+0.60}_{-0.60}) \, 10^{-5}$ &					    & $2.71 \times 10^{29}$ & 0.96 & 7\\\vspace{1mm}

190156.39--365728.4        &14.81       $^{+1.98}_{-1.57}$  & 79.4 $^{+32.1}_{-16.6}$  & 			 & $(1.18^{+0.15}_{-0.15}) \, 10^{-4}$ &					    & $3.80 \times 10^{29}$ & 1.68 & 3\\\vspace{1mm}

190200.11--370222.3        & 0.51       $^{+0.02}_{-0.02}$  &  2.2 $^{+0.04}_{-0.04}$  & 13.2 $^{+1.6}_{-1.5}$    & $(9.87^{+0.68}_{-0.68}) \, 10^{-4}$ & $(2.06^{+0.26}_{-0.26}) \, 10^{-5}$  & $1.10 \times 10^{29}$ & 0.35 & 8\\\vspace{1mm}

190201.92--370743.0        & 0.005      $^{+0.006}_{      }$&  9.4 $^{+0.2}_{-0.2}$    & 22.2 $^{+0.9}_{-0.9}$    & $(4.96^{+0.14}_{-0.14}) \, 10^{-4}$ & $(6.02^{+0.17}_{-0.17}) \, 10^{-4}$  & $2.35 \times 10^{30}$ & 1.30 & 87\\\vspace{1mm}


190211.99--370309.4        & 0.47       $^{+0.07}_{-0.06}$  &  7.9 $^{+0.8}_{-0.8}$    & 			 & $(1.35^{+0.19}_{-0.18}) \, 10^{-5}$ &					  & $2.92 \times 10^{28}$ & 2.36 & 1\\\vspace{1mm}

190222.13--365541.0        & 0.25       $^{+0.01}_{-0.01}$  &  9.5 $^{+0.2}_{-0.2}$    & 25.6 $^{+0.9}_{-1.0}$    & $(2.27^{+0.09}_{-0.09}) \, 10^{-4}$ & $(4.66^{+0.11}_{-0.11}) \, 10^{-4}$  & $1.53 \times 10^{30}$ & 0.96 & 97\\\vspace{1mm}

190227.05--365813.2        & 0.16       $^{+0.04}_{-0.04}$  & 39.7 $^{+6.7}_{-5.0}$    & 			 & $(3.01^{+0.18}_{-0.18}) \, 10^{-5}$ &					   & $7.97 \times 10^{28}$ & 0.71 & 6\\\vspace{1mm}

190233.07--365821.1        & 0  $^{+0.10}_{     }$  & $>740^{    }_{-246}$     & 			 & $(2.22^{+0.29}_{-0.27}) \, 10^{-5}$ &					   & $6.13 \times 10^{28}$ & 1.87 & 3\\
\noalign{\smallskip}\hline														 
\end{tabular}
\end{centering}\\
$^{\bf [1]}$ iron abundance adjusted to fit the 6.7~keV line; \,\,\,
$^{\bf [2]}$ fit including flourescent iron line at 6.4~keV\,\,\, 
\end{table*}


\end{document}